\documentclass[pdflatex,sn-mathphys-num]{sn-jnl}

\usepackage{graphicx}%
\usepackage{multirow}%
\usepackage{amsmath,amssymb,amsfonts}%
\usepackage{amsthm}%
\usepackage{mathrsfs}%
\usepackage[title]{appendix}%
\usepackage{xcolor}%
\usepackage{textcomp}%
\usepackage{manyfoot}%
\usepackage{booktabs}%
\usepackage{algorithm}%
\usepackage{algorithmicx}%
\usepackage{algpseudocode}%
\usepackage{listings}%
\usepackage{subcaption}
\usepackage{float}
\usepackage{natbib}
\usepackage[LGR, T1]{fontenc}
\usepackage[utf8]{inputenc}

\usepackage{tipa}

\newcommand{\CITE}[1]{ \cite{#1}}

\theoremstyle{thmstyleone}%
\theoremstyle{thmstyletwo}%

\theoremstyle{thmstylethree}%

\begin{document}

\title[Analysis of the Maximum Prediction Gain of Short-Term Prediction on Sustained Speech]{Analysis of the Maximum Prediction Gain of Short-Term Prediction on Sustained Speech}


\author*[1]{\fnm{Reemt} \sur{Hinrichs}}\email{hinrichs@tnt.uni-hannover.de}

\author[1]{\fnm{Muhamad} \sur{Fadli Damara}}

\author[2]{\fnm{Stephan} \sur{Preihs}}

\author[1]{\fnm{Jörn} \sur{Ostermann}}

\affil*[1]{\orgdiv{Insitut für Informationsverarbeitung}, \orgname{Leibniz University Hannover}, \orgaddress{\street{Appelstr. 9a}, \city{Hannover}, \postcode{30167}, \state{Lower Saxony}, \country{Germany}}}

\affil[2]{\orgdiv{Institute for Communications Technology}, \orgname{Leibniz University Hannover}, \orgaddress{\street{Appelstr. 9a}, \city{Hannover}, \postcode{30167}, \state{Lower Saxony}, \country{Germany}}}


\abstract{Signal prediction is widely used in, e.g., economic forecasting, echo cancellation and in data compression, particularly in predictive coding of speech and music. 
Predictive coding algorithms reduce the bit-rate required for data transmission or storage by signal prediction. The prediction gain  is a classic measure in applied signal coding of the quality of a predictor, as it links the mean-squared prediction error to the signal-to-quantization-noise of predictive coders. To evaluate predictor models, knowledge about the maximum achievable prediction gain independent of a predictor model is desirable. In this manuscript, Nadaraya-Watson kernel-regression (NWKR) and an information theoretic upper bound are applied to analyze the upper bound of the prediction gain on a newly recorded dataset of sustained speech/phonemes. 
It was found that for unvoiced speech a linear predictor always achieves the maximum prediction gain within at most 0.3 dB. On voiced speech, the optimum one-tap predictor was found to be linear but starting with two taps, the maximum achievable prediction gain was found to be about 2 dB to 6 dB above the prediction gain of the linear predictor. Significant differences between speakers/subjects were observed.

The created dataset as well as the code can be obtained for research purpose upon request.}

\keywords{prediction gain, mutual information, conditional expectation, speech coding, kernel regression, speech prediction
}



\maketitle

\section{Introduction}
Signal prediction is widely used in, e.g., economic forecasting, echo cancellation, and data compression, particularly in predictive coding of speech and music\CITE{Economicforecasting,Ivry21, backstrom2017speech}. 
Predictive coding algorithms reduce the bit-rate required for data transmission or storage by signal prediction.

In predictive coding, a source signal is compressed by prediction and subsequent quantization of the prediction residual. As coding performance is strongly affected by the prediction error respectively the prediction error variance\CITE{chu2004}, extensive research in earlier decades focused on better predictor models aiming to improve simple linear predictors by considering non-linear models like volterra filters or neural networks\CITE{thyssen1994non,Birgmeier96,Volterra, Zhao18}.
A thorough review on predictive coding can be found in\CITE{Sheferaw23}.
\subsection{The Prediction Problem}
Prediction of a signal\footnote{Depending on context, $y(n)$ is either the entire signal or the signal at time $n$.} $y(n)$ is the application of a function $f$, commonly referred to as a predictor,  that uses past signal samples $\mathbf{y}(n):= (y(n), y(n-1), \dots, y(n-(P-1)))$ to estimate future signal samples $y(n+L)$ with $\mathbb{N}\ni L\geq 1$. $L$ denotes the number of time or sample steps to be predicted ahead, and $P$, the so called "tap size", is the number of considered previous samples.
Prediction with $L=1$ is referred to as one-step prediction, sometimes called short-term prediction. 
The main goal in signal prediction is to find a predictor that minimizes some distance measure, most commonly the mean-squared (prediction) error $E((y(n+L)-f(\mathbf{y}(n)))^2)$, which is also the prediction error variance. The error minimizing predictor is also called the optimal predictor (with respect to the chosen distance measure).
The prediction problem, in the case of the mean-squared error, can thus be formalized as 
\begin{equation}
    \underset{f\in \mathcal{F}}{inf} E((y(n+L) - f(\mathbf{y}(n))^2)
\end{equation}
with some function space $\mathcal{F}$.

Generally, the optimal predictor $f_{opt}$ with respect to the mean-squared error is the conditional expectation $f_{opt}(\mathbf{y}(n)) \equiv E[y(n+L)|\mathbf{y}(n)]$\CITE{boxjenkins}. However, in practice this function or its corresponding mean-squared error is difficult to obtain. 

Linear prediction is an approximate approach to the optimal prediction problem aiming to find the optimal predictor $f$ within the class of linear functions (compared to the very general space  of, e.g., continuous functions). The optimal solution can be found using the Wiener-Hopf equation \cite{WienerHopf}.
Nonlinear prediction usually attempts to improve on linear prediction by considering a broader search space of functions, e.g., the space of polynomials of some maximum degree as in Volterra filters\CITE{Volterra}.

Knowledge of the minimum achievable prediction error variance for a given type of signal is generally desirable. If a predictor achieves such a minimum, no further improvement of the predictor is possible, and future developments could focus on other parts of, e.g., a coding algorithm, for example the quantization and entropy coding stage.

\subsection{Background}
While current state-of-the-art audio codecs are mostly based on neural networks, which do not perform explicit predictions but usually learn a data representation to compress audio signals on a window-by-window basis at very low bit-rates \CITE{Soundstream, Encodec}, their algorithmic latency typically exceeds 10 ms.
Predictive coders can achieve algorithmic latencies considerably lower than 10 ms -- using backward adaptive prediction, an algorithmic latency of zero can be achieved \CITE{Gayer04}. Such low latencies can be desirable for hearing aid applications \CITE{Hinrichs19,Hinrichs21} as well as certain live music applications \CITE{Preihs15}. 

Additionally, a branch of neural audio codecs merges traditional linear predictive coding (LPC) methods with neural networks\CITE{LPCNet2, LPCNet}, where LPC is used to compute predictions that are further processed by recurrent networks. Such approaches could potentially benefit from improved predictor models.
As such, investigations on prediction algorithms can still be of interest.

For linear prediction, a lower bound of the mean-squared (prediction) error (MSE) can be derived using the power spectral density of a signal\CITE{szego1958}. By comparing the spectral flatness of a given signal with that of a white noise signal, the maximum achievable prediction gain  can be calculated.  The prediction gain is the logarithmic ratio of signal variance and prediction error variance to be defined in later parts of the manuscript.
Using the finite past of a signal, nonlinear predictors can often achieve substantially lower MSE than linear predictors, and Kanter \CITE{kanter1979lower} and Shepp et al.\CITE{Slepian} derived a formula for the maximum possible improvement of a nonlinear predictor over a linear one for stationary moving average processes. 

To the best knowledge of the authors, specifically for speech signals, the only publications that are at least partially concerned with estimating the maximum achievable prediction gain (see Eq. \ref{eq:PG}) of speech are\CITE{Birgmeier97,bernhard1998tight} and\CITE{bernhardPhD}.

Most importantly for this work, Bernhard\CITE{bernhard1998tight} derived an upper bound for the prediction gain of a linear or non-linear predictor based be obtained through the automutual information function. 
In order to estimate this upper bound according to\CITE{bernhard1998tight}, a signal has to be strict-sense stationary\footnote{A strict-sense stationary signal or process has time-invariant distributions, whereas a wide-sense stationary process only requires a time-invariant autocorrelation (or covariance) function (as well as constant mean and finite variance)\CITE{park2018fundamentals}}. 
Although a speech signal is usually  considered stationary across periods of around $30\,\text{ms}$\CITE{Paliwal10}, the obtainable sample size during such a time frame is insufficient to apply the upper bound formula given in \cite{bernhard1998tight} because of the estimations involved. Longer durations of stationary speech can be achieved by sustaining segments of speech, which refers to the continuous production of isolated phonemes. This approach was used in\CITE{bernhardPhD} to investigate predictor models, but neither was the stationarity condition checked nor detailed information about the maximum prediction gain with respect to tap size, phoneme, etc. reported. Additionally, key properties like bias and variance of the utilized mutual information estimator were  assessed only for the bivariate Gaussian case (bias was also assessed for a harmonic superposition).

\subsection{Contribution}
In this work, an analysis of the maximum achievable prediction gain on sustained phonemes is presented. 
Two estimators of the mutual information are first evaluated on artificial data to assess their bias and variance.
These estimators are then applied to estimate the upper bound of the prediction gain on sustained voiced and unvoiced speech recordings of five subjects. By comparing static with adaptive linear prediction, the instationarity of the speech signals is assessed. Furthermore, using kernel regression, the conditional expectation is estimated and a second upper bound calculated to supplement the results of the mutual information based upper bound.

\subsection{Overview}
Section \ref{ssec:upperbound} introduces the previously mentioned upper bound of the prediction gain derived  in\CITE{bernhard1998tight}.
It is based on two main parts, the automutual information function of a random process and the entropy difference between a random process of interest and a Gaussian random process of equal variance.
Subsequently,  two estimators of the mutual information, used in this work to estimate both the automutual information and entropy of speech signals, are described.
This section is followed by the description of the method to assess the key  characteristics bias and variance of these estimators. Such an evaluation is necessary to be able to gauge the  estimation error on speech data.
For this purpose, signal models are needed with analytic expressions for the automutual information.  The automutual information function is known for Gaussian processes which made them a prime choice for these investigations.

Another way to estimate the maximum prediction gain is to estimate the conditional expectation, and to use it to compute the optimal prediction error (in terms of error variance). Comparing this optimal prediction error variance with the variance of the respective speech signal allows to compute the maximum prediction gain.

This is achieved using Nadaraya-Watson kernel regression, described together with the optimization involved in Section \ref{ssec:kernel}.

All of these involved estimators require a substantial amount of samples from stationary speech signals. As there was no publicly available dataset, a custom dataset was recorded. The recording of these speech signals is explained in Section \ref{ssec:speechmaterial}. The method used to measure the stationarity of the speech signals is described in Section \ref{ssec:stationarity}

In the evaluations, these different ways to estimate the maximum prediction gain are compared to several predictor models. A linear predictor is applied to be able to assess the maximum benefit of general nonlinear prediction. Two nonlinear predictors, both neural networks, are used to compare their prediction gain to the estimated upper bound(s).

Finally, Section \ref{sec:results} presents the results of the bias and variance evaluations of the mutual information estimators, followed by the main result, the evaluation of the maximum prediction gain on sustained speech. These results are discussed in Section \ref{sec:discussion}.

\section{Methods and Materials}

\subsection{An Upper Bound of the Prediction Gain:}
\label{ssec:upperbound}
The prediction gain is defined as
\begin{align}
\label{eq:PG}
G = 10 \log_{10} \frac{\sigma_y^2}{\sigma_{e}^2},
\end{align}
where $\sigma_y^2$ is the variance of the signal $y(n)$, and $\sigma^2_{e}$ is the variance of the  prediction error of a given predictor. 

An  upper bound of the (one-step) prediction gain, independent of a specific predictor model, is given\CITE{bernhard1998tight} by
\begin{align}
\label{eq:maxPG}
G \leq G_{\text{max}} = 6.02\cdot(I(y(n+1);\mathbf{y}(n)) + \Delta),
\end{align}
consisting of the so called automutual information function\CITE{bernhard1998tight} (in this work $L = 1$),
\begin{equation}
I(y(n+L);\mathbf{y}(n)) \equiv I(y(n+L);y(n), y(n-1), \ldots, y(n-(P-1)))
\end{equation}
and the entropy difference
\begin{equation}
\label{eq:delta}
\Delta = H_{gauss}(\sigma_y^2) - H(y(n)).
\end{equation}
$I(y(n+L);\mathbf{y}(n))$ measures the amount of information contained in the past of $y(n)$ about $y(n+L)$. $\Delta$ is the difference between the  differential entropy $H_{gauss}(\sigma_y^2) = \frac{1}{2} \log_2 (2 \pi e \sigma_y^2)$ of a Gaussian random variable with the same variance $\sigma_y^2$ as $y(n)$. $H(y(n))$ is the (unconditional) differential entropy of the signal $y(n)$. 
For two sets of continuous random variables $X^n$ and $Y^n$ with $X^n:=\lbrace X_1, X_2, \ldots, X_n \rbrace$, $Y^n:=\lbrace Y_1, Y_2, \ldots, Y_n \rbrace$ the mutual information is defined\CITE{cover2012elements} as
\begin{align}
I(X^n;Y^n) = \int_\mathcal{S} p(x^n,y^n) \log_2 \left( \frac{p(x^n,y^n)}{p(x^n) p(y^n)} \right) dx^n dy^n,
\end{align}
where $p(x^n,y^n)$ is the joint probability density function with support  $\mathcal{X}\times\mathcal{Y}=\mathcal{S}$. The differential entropy $H(X^n)$ of a set of continuous random variable $X^n$ is defined \CITE{cover2012elements} as
\begin{align}
H(X^n) = -\int_\mathcal{X} p(x^n) \log_2 p(x^n) dx^n
\end{align}
where $\mathcal{X}$ is the support of the density function $p(x^n)$. 
The upper bound according to Eq. \ref{eq:maxPG} will be referred to as maximum prediction gain estimate (PGMAXE).

\subsection{Mutual Information Estimation:}
The Kraskov-St\"ogbauer-Grassberger (KSG) \CITE{kraskov2004estimating} estimator and the Fast Mutual Information (FMI) estimator\CITE{bernhard1994fmi} were used to obtain the PGMAXE.
The KSG is a well known and good performing algorithm for estimating mutual information. The FMI was used in the reference work in\CITE{bernhardPhD} and part of the original code received from the author. As such it was natural to compare the FMI to the KSG estimation which has not been done before.

Both estimators use a single parameter, $k$ and $\chi_{\alpha}$ for the KSG and FMI, respectively, which controls the estimation trade-off between bias and variance. For the KSG, $k=2$, for minimal bias, was used throughout all experiments. For the FMI, $\chi_{\alpha} = 2$, for minimal bias, was used throughout all experiments. To estimate $\Delta$ of Eq. \ref{eq:delta}, the respective mutual information estimators and the method suggested by Bernhard \CITE{bernhard1998tight} were used. 
For this purpose, Bernhard provided his original code which was used for any analyses concerning the FMI estimator.

\subsection{Evaluation of the Mutual Information Estimators:}
Gaussian AR($N$) processes were employed to investigate the bias and variance of the mutual information estimators.
The bias $B(\theta, \hat{\theta})$ of an estimator $\hat{\theta}$ for a parameter $\theta$ is defined as 
\begin{equation}
    B(\theta, \hat{\theta}) := E(\hat{\theta}) - \theta
\end{equation}
and the variance $Var(\hat{\theta})$ as 
\begin{equation}
    Var(\hat{\theta}) := E((\hat{\theta}-E(\hat{\theta}))^2).
\end{equation}
Bias measures the distance between the average value of many estimations of a parameter and its ground truth value. Variance is a measure of the dispersion of the estimates around the bias.

The AR($N$) parameters were extracted from real speech recordings of voiced and unvoiced speech and for every process, 50 realizations were generated. The numbers of samples were varied between $10^2$ and $10^6$ samples. While the main results in this work use 10k samples and between one to three taps, as a reference, the estimator performance is reported for larger and smaller sample sizes as well as a larger number of AR coefficients. The analytic mutual information of a Gaussian AR($N$) is given\CITE{pinsker1964information} by 
\begin{align}
I_{\text{ana}}(X;Y^n) = -\dfrac{1}{2} \log_2 \left( \dfrac{\sigma_X^2 \det(K_{Y^n})}{\det(K_{XY^n})} \right),
\label{eq:Igauss}
\end{align}
where $\det(K_{Y^n})$ is the determinant of the covariance matrix $K_{Y^n}:=E[(y_i - E[y_i]) (y_j - E[y_j])^T]$ of the set of random variables $Y^n$ and $\det(K_{XY^n})$ is the determinant of $K_{XY^n}$, the analogously defined covariance matrix but including the random variable $X$. The bias was then calculated by $\frac{1}{M} \sum_{i=1}^M (I_{\text{ana}} - \hat{I}_i)$, where $M$ is the number of realizations which was equal to 50. The variance was calculated using the standard unbiased estimator $\frac{1}{M-1}\sum_{i=1}^{M}(\hat{I}_i - \overline{\hat{I}_i})^2$. $\overline{\hat{I}_i}$ is the sample mean of the mutual information estimate across the $M$ repetitions.

Further investigations were performed regarding the impact of minor instationarity of a signal on the PGMAXE on AR(1) and AR(2) processes revealed that the PGMAXE remains approximately correct ($\pm 0.2$ dB) for linear and sine-like variation of the excitation noise power. For example, for the sine-like variations, the noise power $\varepsilon_n$ driving the processes was modified according to $\hat{\varepsilon}_n:= \varepsilon (1 + A sin(2\pi f n))$ with small constants $A$ and $f$.
Sine-like variation of AR(2) poles showed that the PGMAXE remained approx. correct when the pole magnitudes were varied by $\pm 5\, \%$.
\begin{figure}
    \centering
    \subfloat[]{\includegraphics*[angle=-90,width=0.45\linewidth]{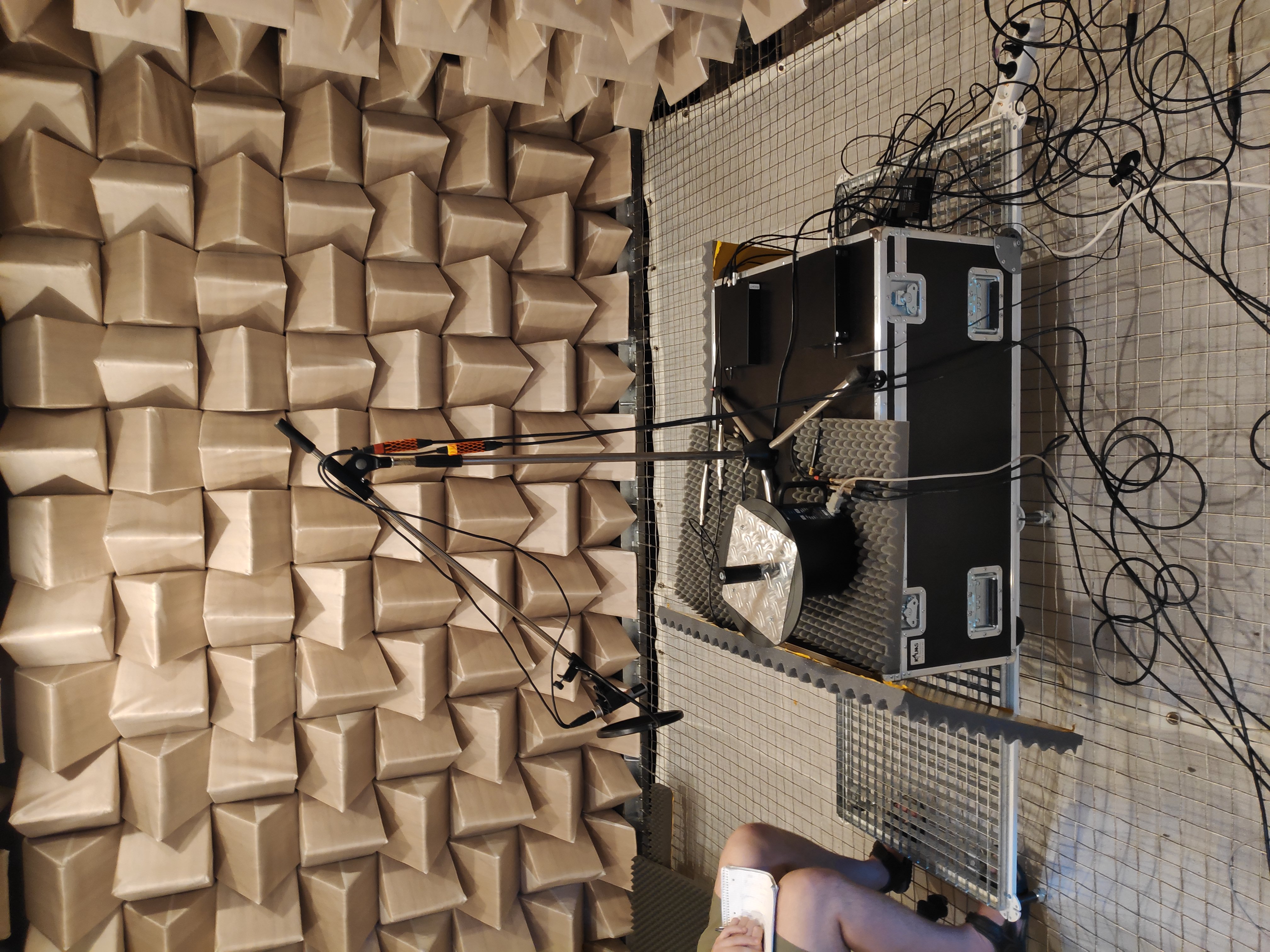}}\hfill
    \subfloat[]{\includegraphics*[angle=-90,width=0.45\linewidth]{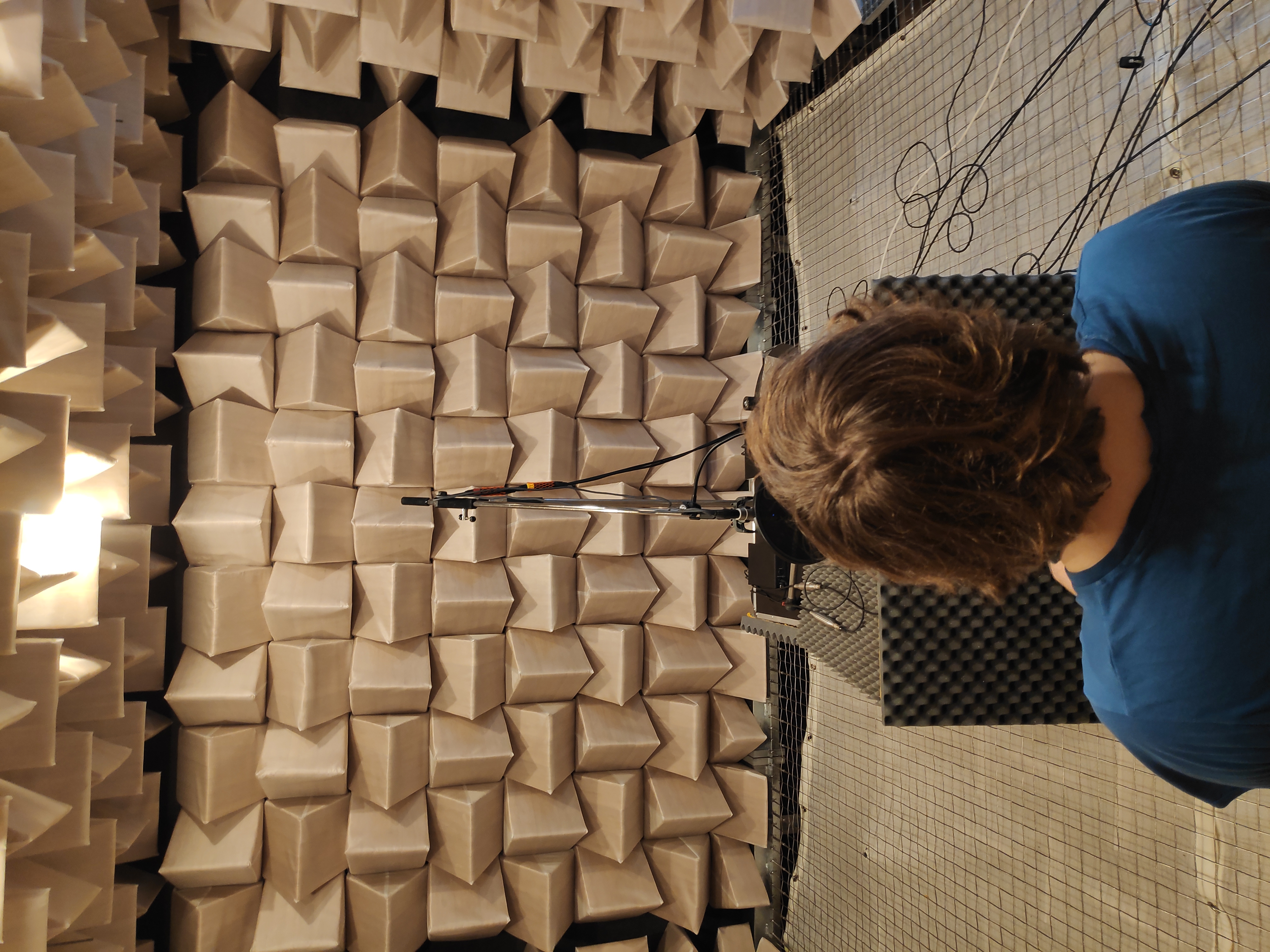}}
    \caption{Setup inside the anechoic chamber. A chair was positioned right infront of a microphone. Unlike depicted, for recording, the microphone was positioned closer to the mouth of the subjects.}
    \label{fig:setup}
\end{figure}
\subsection{Estimation of the Conditional Expectation:}
\label{ssec:kernel}
To estimate the conditional expectation $E(Y_n|Y_{n-1}, \dots, Y_{n-P})$, multivariate  Nadaraya-Watson kernel regression (NWKR) with Gaussian kernels\CITE{KernelRegression} was applied. The NWKR estimates the conditional expectation using
\begin{equation}
    \hat{E}(Y|Y=x) = \frac{\sum_{i=1}^{T} K_h(x-x_i)y_i}{\sum_{i=1}^{T} K_h(x-x_i)},
    \label{eq:NWKR}
\end{equation}
where $K_h(x):= \frac{1}{h}e^{-\frac{x^2}{2 h^2}}$ is the kernel, $x_i,y_i$ are the sampled data points, and $T$ is the total number of samples. The conditional expectation according to Eq. \ref{eq:NWKR} is therefore estimated by a locally weighted average of the sample points. $h\in \mathbb{R}^P$ denotes the kernel's bandwidth, acting as a stardard deviation, whose value is crucially affecting the estimation. Through leave-one-out cross-validation (LOOCV)\CITE{ALTMAN1995195}, the optimal bandwidths were obtained per audio recording/segment, maximizing the prediction gain using Matlab's \textit{fmincon} function. Five random starts in the interval $[0.0001, 0.1]$ were performed and the best optimization result with respect to the PG was used.
Our LOOCV implementation was validated on Gaussian autoregressive processes of order $N\in\{1,5,9\}$, using random poles between $0.3$ and $0.999$, as well as selected speech data. The Gaussian autoregressive processes allowed to easily vary the strength and complexity of the stochastic dependency and thus were a natural choice.

Additional evaluations were performed in\CITE{Hinrichs24} on processes of the kind of $y(n) = k\cdot e^{-|y(n-1)|} + \epsilon(n)$ with unit variance zero mean Gaussian white noise $\epsilon(n)$, where the optimal predictor is non-linear. The exponential function allows to easily achieve a stable process while introducing nonlinear dependencies.

It was confirmed that the NWKR does not undersmooth using the LOOCV method~-- i.e., does not yield very small bandwidths resulting in delta-function like kernel values --  and converges to reasonable bandwidths in any case. It was observed that the NWKR either achieved the maximum PG or achieved slightly suboptimal PG, even when the bandwidths' were initialized with very small values (e.g. $10^{-7}$), where a direct optimization without LOOCV resulted in significant undersmoothing every time.

\subsection{Speech Material}
\label{ssec:speechmaterial}
All recordings were performed with a RME Fireface UC audio interface at +35 dB analog gain using a Sennheiser MKH800 Twin microphone and the Reaper DAW in an anechoic chamber at a sampling rate of 48 kHz with 24 bits/sample. The setup inside the anechoic chamber is shown in Fig. \ref{fig:setup}. 

Only the channel of the microphone facing the subjects was used. Five subjects, labeled S1 to S5, were recorded of which three were male (S1 to S3) and two were female (S4 and S5) in the age range 21 to 32 years. Each subject was asked to sustain the voiced German vowels and consonants /a:/, /e:/, /i:/, /o:/, /u:/, /n/ and /\textipa{\ng}/ as well as the unvoiced consonants /f/, /s/ and /\textipa{S}/.  
Ten repetitions  were recorded for each phoneme. Each subject was instructed to move their body and head as little as possible during the speech production to achieve maximum stationarity.
\begin{figure*}
    \centering
    \subfloat[]{\includegraphics*[width=0.49\linewidth]{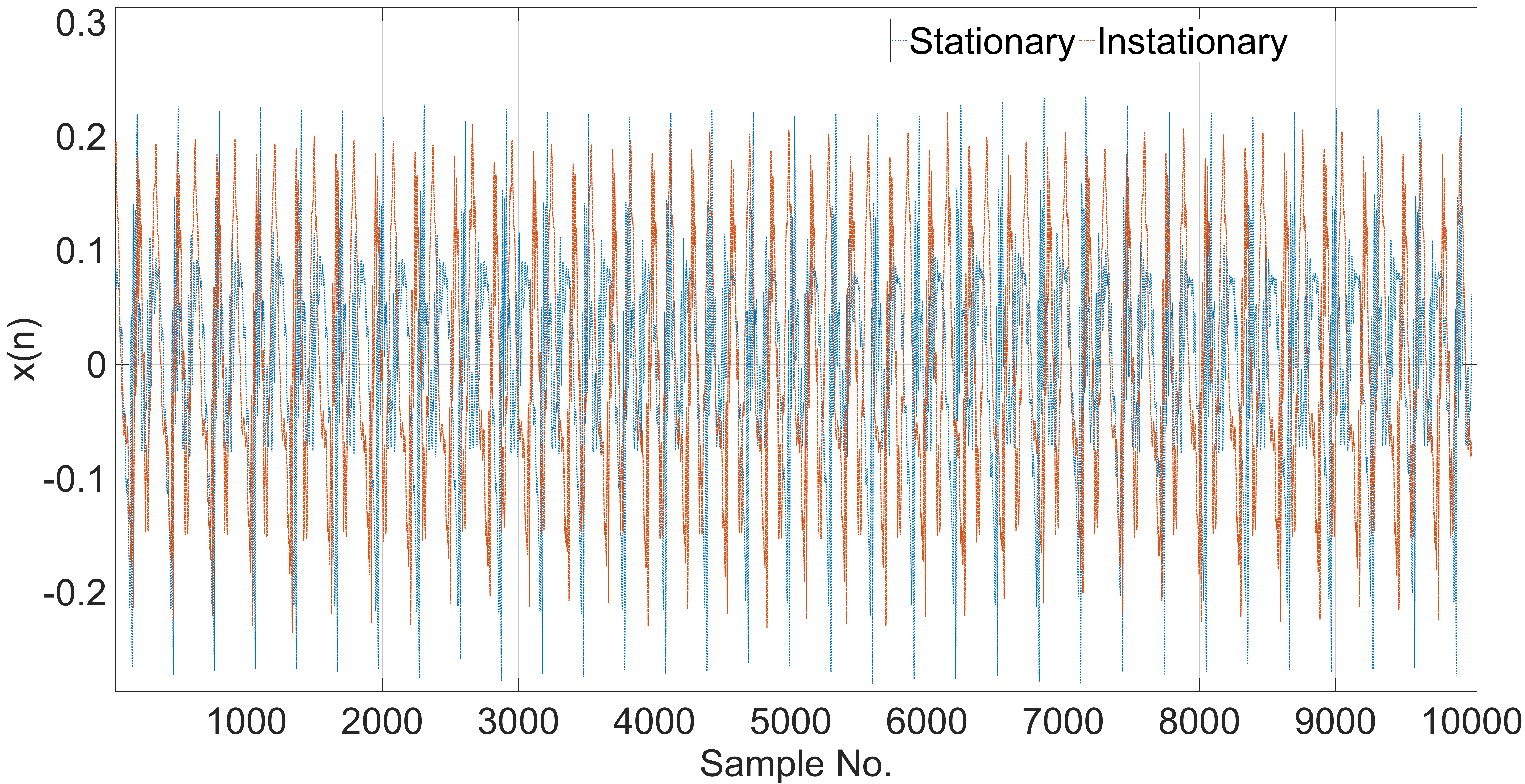}}\hfill
    \subfloat[]{\includegraphics*[width=0.49\linewidth]{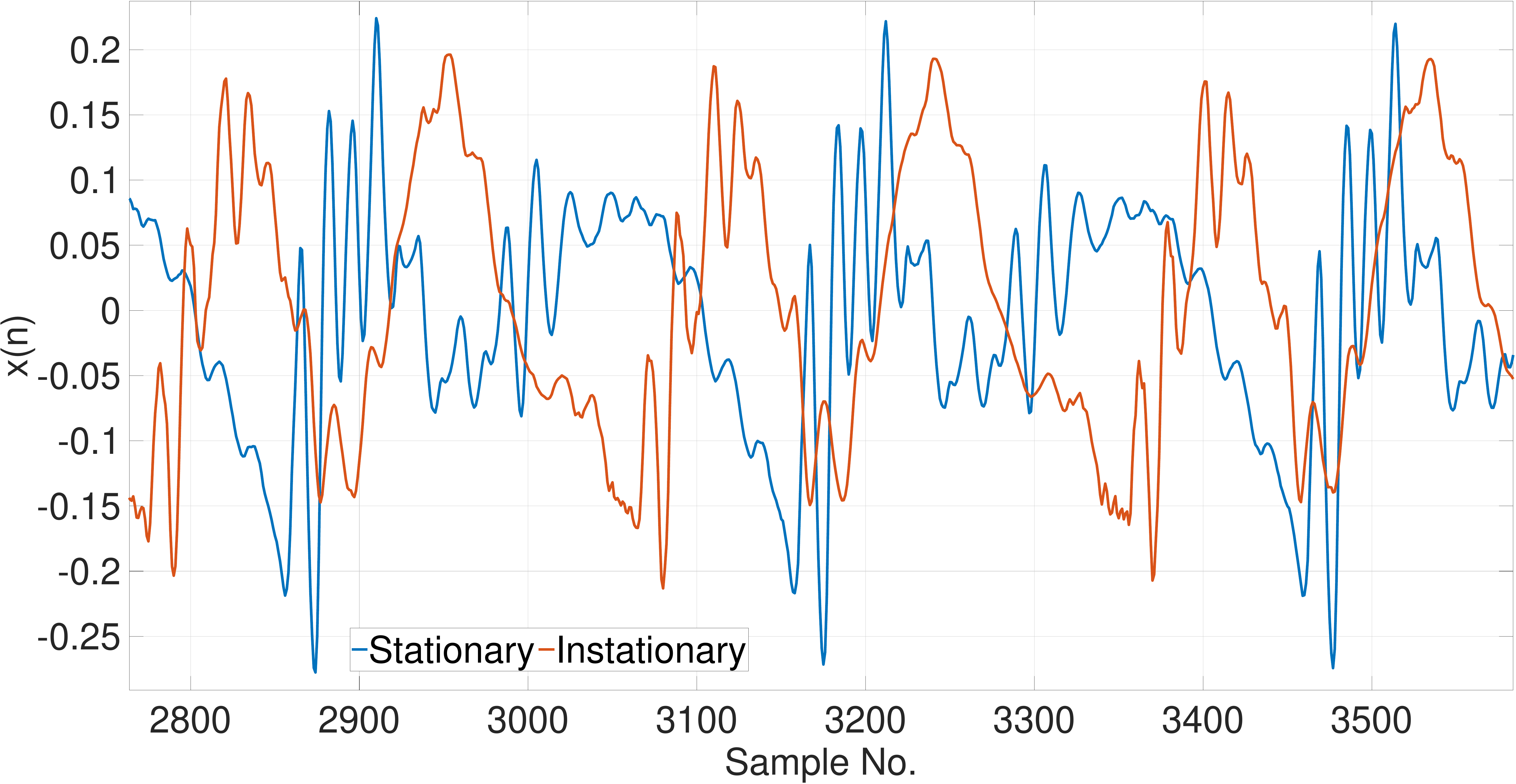}}
    \caption{Example waveforms from the 10,000 samples long audio snippets. The samples were taken from two different subjects, each producing a sustained /e:/ vowel. (a) shows the entire sequence and (b) a smaller segment to highlight the finer structure. The waveform deemed stationary exhibited a difference of at most 0.13 dB in prediction gain between the static and the adaptive linear predictor, where this maximum was achieved for the two step predictor. The waveform deemed instationary exhibited a difference of at most 1.17 dB in prediction gain. This maximum was achieved for the one step predictor. Some minor variations can still be observed in both waveforms.}
    \label{fig:waveforms}
\end{figure*}
As it was found that a comfortable posture was most important for consistent speech production, the subjects were asked to lean back or forward (before recording) according to comfort. This resulted in varying distances of the speakers mouths and the microphone position.
The raw duration of each phoneme ranged from about 5 to 15 seconds, corresponding to between 240,000 and 720,000 samples. From these, the initial transition and the ending fade out, in total around a half second, was cut and the remainder of the recording was then split into sections of 10,000 samples, referred to in the following as segments.
Longer segments -- 50k and 100k samples -- were also considered, but the number of stationary segments decreased too much for longer durations to seem viable.
These segments were then independently analyzed. 
Only stationary segments according to the definition given in the following section were considered in the evaluations. Two example waveforms, one deemed stationary, one deemed instationary, are shown in Fig. \ref{fig:waveforms} as a whole and in close up.
Some phonemes were more difficult to sustain than others and for the /n/ phoneme no subject was able to produce even a single stationary segment, defined in Section \ref{ssec:pred_model}, of a length of 10,000 samples and for /o/ only one subject managed to do so. The /e/ phoneme appeared to be the easiest to sustain with respect to stationarity.

\subsection{Measuring Stationarity}
\label{ssec:stationarity}
A sample-adaptive linear predictor (SALP) was applied as an ad-hoc measure of the stationarity of the segments.
The idea is that a linear predictor, adapted sample by sample, should achieve a larger prediction gain than a static linear predictor if the process to be predicted is instationary. The more pronounced the instationarity, the greater the difference in prediction gain should be.

The coefficients of the SALP were adapted at every step with an optimum buffer size which was selected from $\{2 \cdot N, 2 \cdot N+1,\dots, 5 \cdot N \} \cup \{ 5\cdot N+1, 5\cdot N+21, \dots, 800\}$, where $N$ is the number of taps of the SALP, such that the PG of the SALP was maximized across the respective recording. The stepsize of $20$ for the search for the optimal buffer size was used to speed up the computation and based on previous experiments, where the optimal buffer size in case of significant instationarity was close to $2\cdot N$ and in all other cases the precise buffer size did not greatly affect the achieved PG. The performance of the SALP-method showed good agreement with the subjective impression of the recordings. Recordings or segments with noticeable changes of the phonemes showed a PG improvement of the SALP compared to the static predictor of between roughly 1 dB and 5 dB, in good agreement with values found in the literature. Little to no improvement on the other hand was observed for segments that were subjectively also rated stationary. Only recordings where the SALP with one, two as well as three taps achieved an improvement of the PG of less than 0.2 dB compared to the static linear predictor were considered sufficiently stationary. All predictors were trained on the data they were tested on as the number of their model parameters was about two orders of magnitude smaller than the number of training samples.
Note that this way of assessing stationarity can only assess wide-sense stationarity as the linear predictor's computation relies on signal correlation and as such might miss non-linear changes in the dependency structure.
While time-variances in joint densities which are not represented in the correlation can be mathematically conceived, it is assumed that in practice this does not happen, i.e., that a substantial amount of potential change over time in the joint densities is represented in a corresponding change in signal correlation.

Furthermore, our method can only account for short-term time-variance due to using a linear predictor of at most order three, corresponding to the maximum number of predictor taps considered in this work. Long term changes over longer periods cannot be captured this way and appear to be present in general (e.g. slight variations in peak amplitude). However, this should not matter as we are concerned solely with short-term prediction.

Note that a highly similar method of measuring stationarity is discussed in \cite{Haykin14} regarding the "degree of stationarity" of a random process.

We also considered using standard stationarity tests like the Kwiatkowski–Phillips–Schmidt–Shin\CITE{kwiatkowski1992testing} and augmented Dickey-Fuller test\CITE{dickey1979distribution}, but they failed to yield reasonable results. According to these tests, segments were deemed stationary even when the adaptive linear predictor was able to improve on the static one by 1 dB or more in terms of prediction gain.
As such, we proceeded with the proposed SALP-based method.

\subsection{Predictor Models:}
\label{ssec:pred_model}
A static linear predictor, obtained using the covariance method\CITE{Makhoul75}, a radial basis function network (RBF) as in\CITE{Birgmeier96} with 80 neurons in the hidden layer and the bias set to zero as well as a time-delay neural network (TDNN) as presented in\CITE{thyssen1994non} were evaluated in this work.
The purpose of these non-linear models was to serve as a point of reference of the achieved prediction gain by practical predictors. The TDNN was chosen due to historic interest as the -- to the best knowledge of the authors -- first application of neural networks to speech prediction. The TDNN achieved substantial improvements over linear predictors in\CITE{thyssen1994non}, albeit using ten taps compared to at most three in this work.
A RBF network, which is more complex than the TDNN, was also used in the reference work\CITE{bernhardPhD}.
\begin{figure*}[h!] 
\centering                           
\subfloat[KSG Voiced AR(N)\label{fig:KSG_voiced}]{\includegraphics[width=0.49\textwidth]{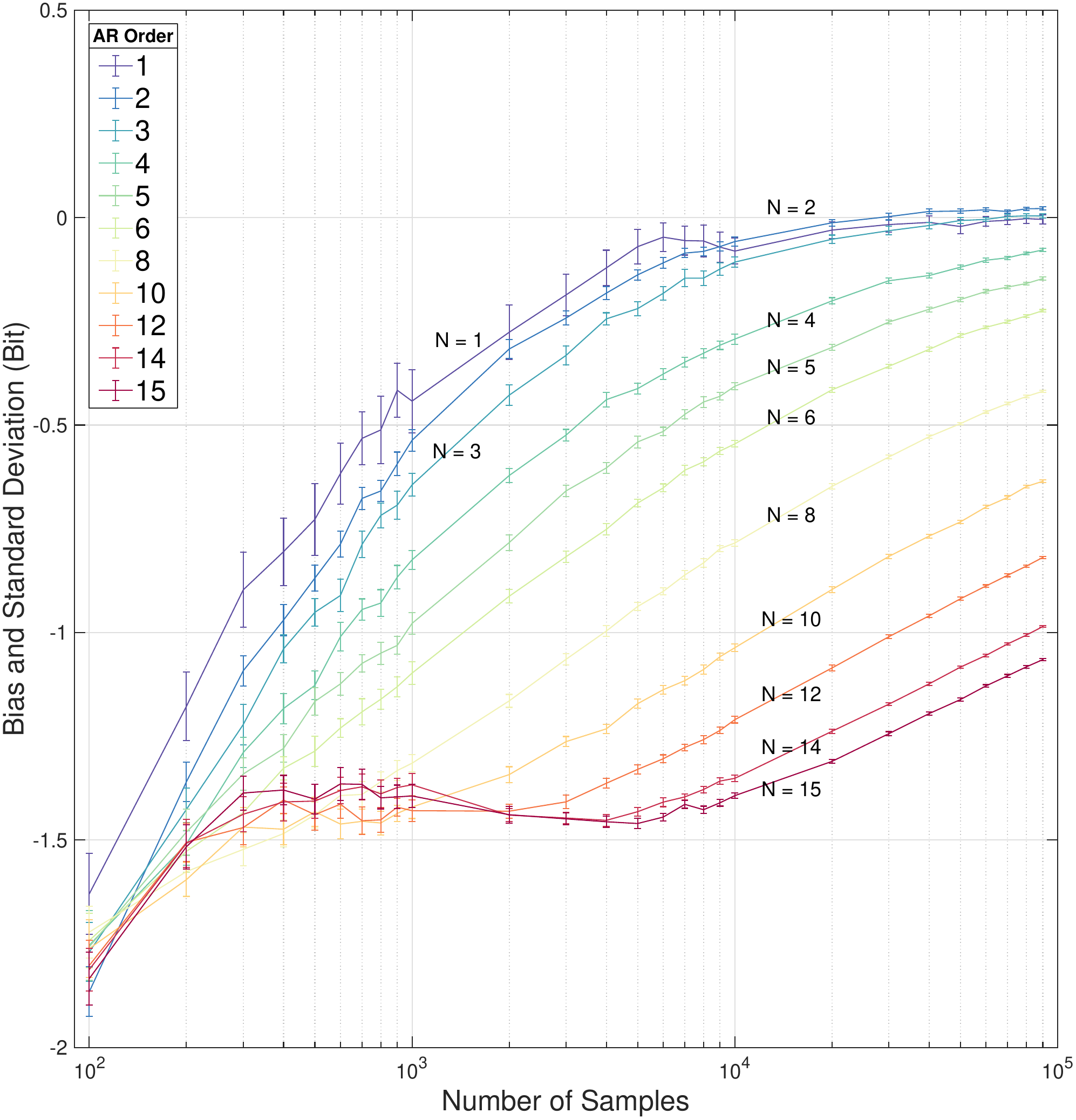}}
\subfloat[FMI Voiced AR(N)]{\includegraphics[width=0.49\textwidth]{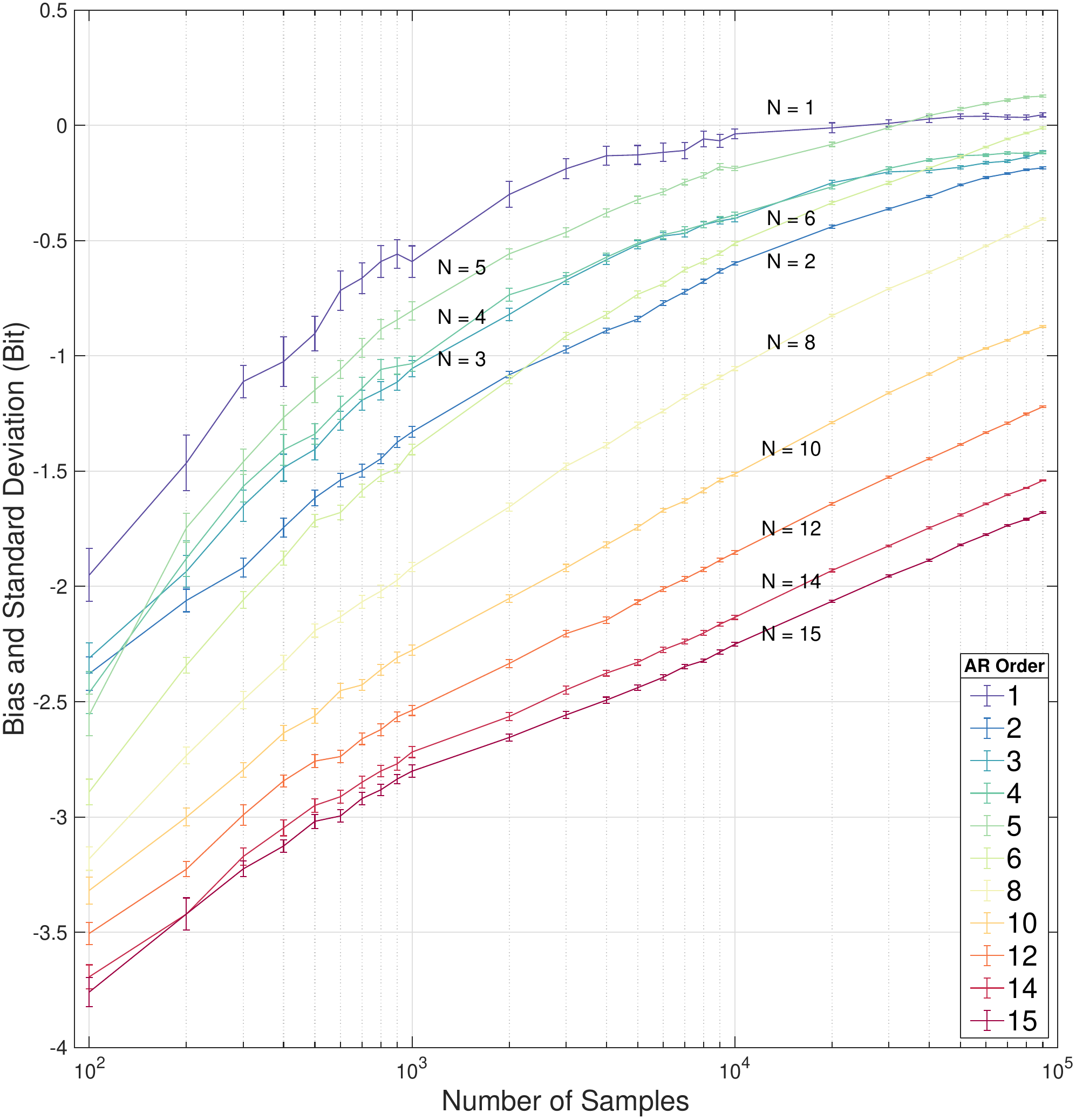}}  \\
\subfloat[KSG Unvoiced AR(N)]{\includegraphics[width=0.49\textwidth]{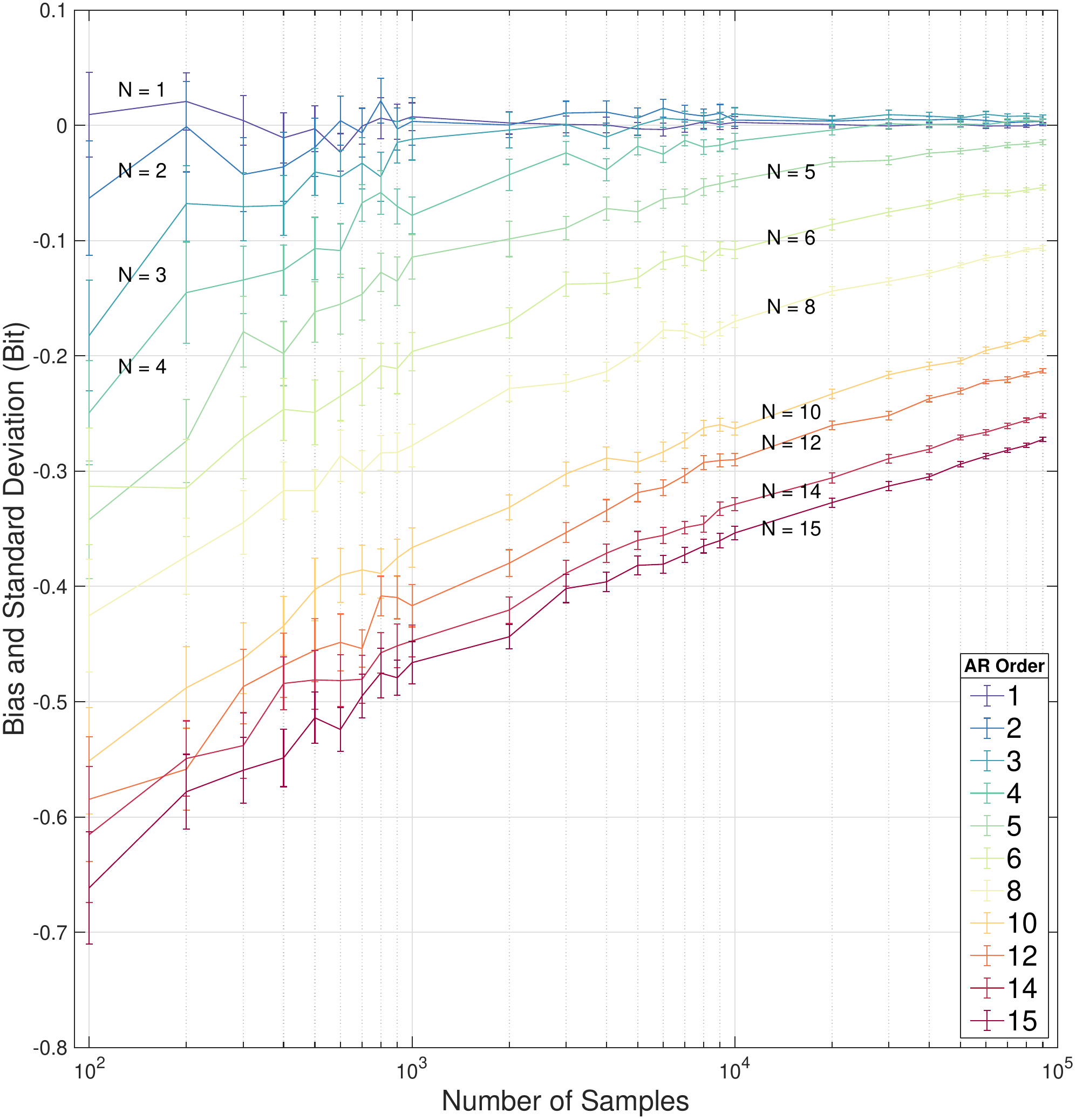}}
\subfloat[FMI Unvoiced AR(N) \label{fig:FMI_unvoiced}]{\includegraphics[width=0.49\textwidth]{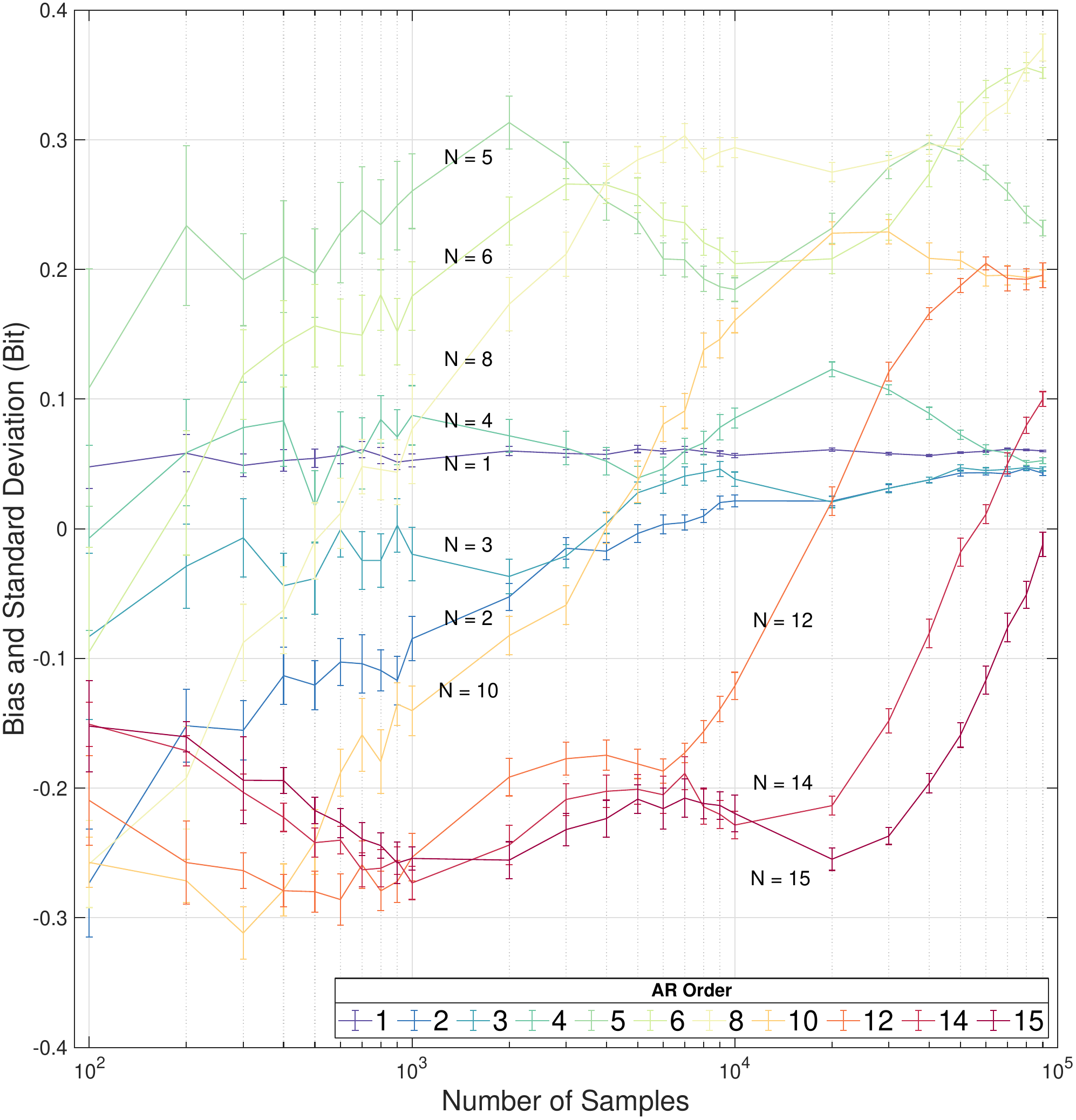}} 
\caption{\label{fig:voiced-a} Estimation bias and standard deviation for the (a) KSG and (b) FMI estimator on AR(N) processes. Process parameters were extracted from the recording of an /a:/ (voiced) and /s/ (unvoiced) phoneme. Error bars indicate $0.25\cdot \sigma$.  
 }
\end{figure*}
While the TDNN, and to a lesser degree the RBF  network are not state-of-the-art models for time series forecasting, which would encompass architectures like long short-term memory networks (LSTMs), gated recurrent units (GRUs) or echo state networks\CITE{Zhao18} -- a type of recurrent network with sparsely connected hidden layers --, these models usually come with a considerably larger number of parameters. This makes it difficult to avoid considerable overfitting given the comparatively short audio recordings, covering only 10,000 samples. However, due to their special production nature, should be used for training and testing. Furthermore, constraining the information of LSTMs and similar networks to the previous K samples appears to be contrary to their design. As such, they were considered unsuitable for the research performed.
\begin{figure*}[b]
\minipage{0.49\textwidth}
  \includegraphics[width=\linewidth]{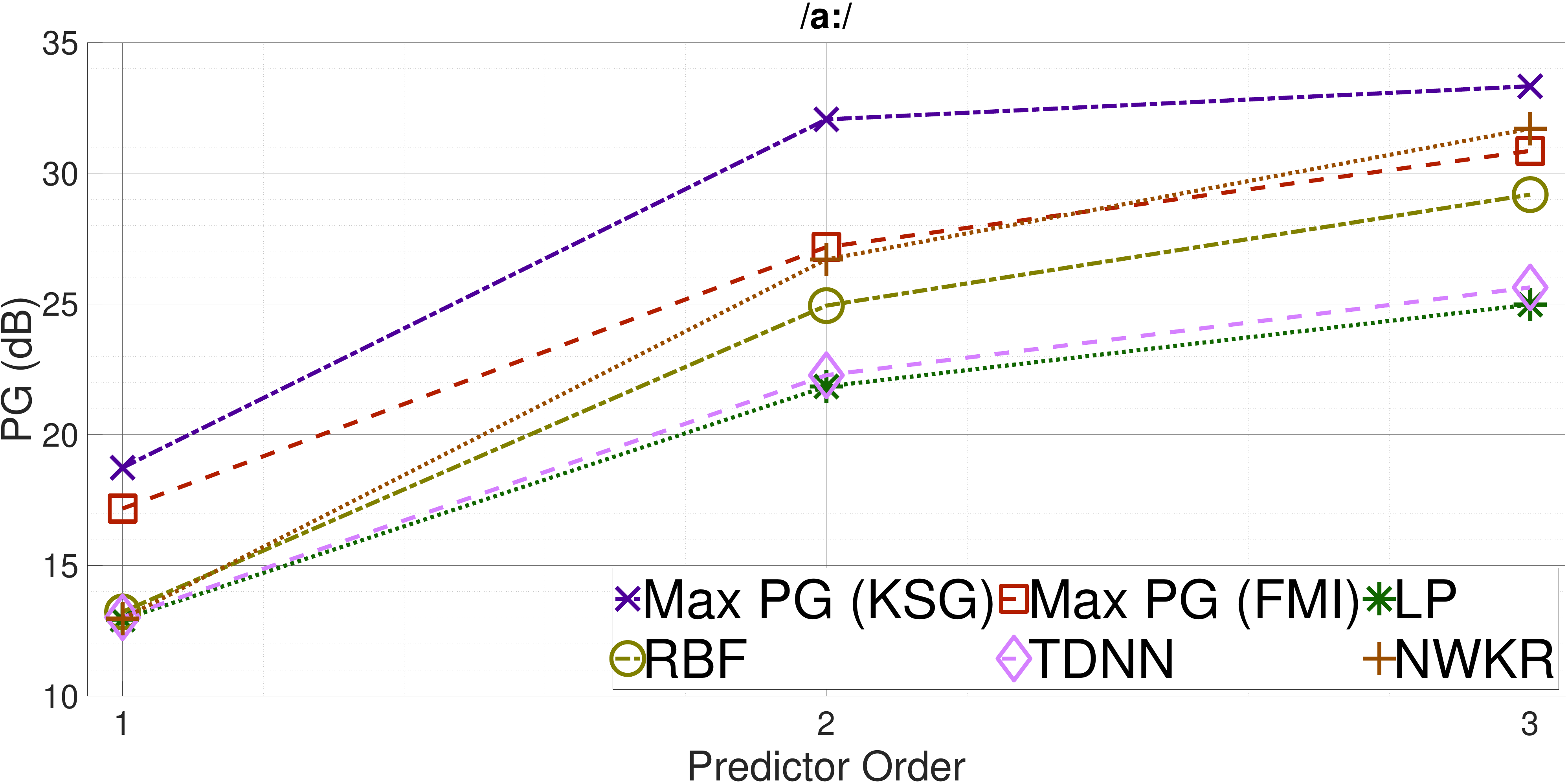}
\endminipage
\minipage{0.49\textwidth}%
  \includegraphics[width=\linewidth]{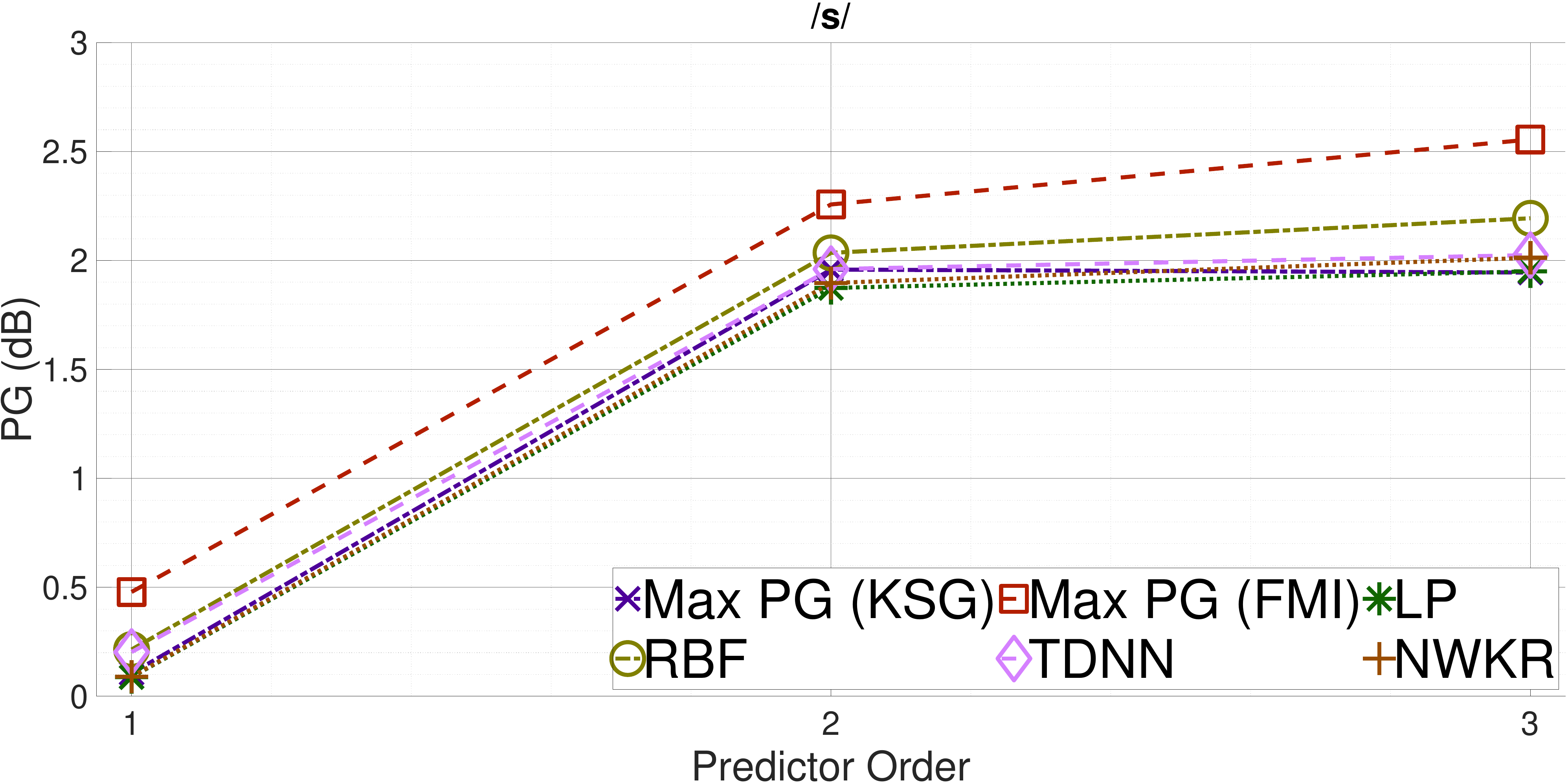}
\endminipage
  \caption{Prediction gain of linear and nonlinear predictors for the phonemes /a:/ (left) and /s/ (right), each corresponding to one segment of a male subject. }\label{fig:three_phonemes}
\end{figure*}
The TDNN and RBF networks were trained ten times for 100 epochs using Levenberg-Marquardt backpropagation\CITE{LevenbergM} with a mean-square error loss. After every training, the prediction gain was assessed to evaluate their  performance.

\section{Results}
\label{sec:results}

\subsection{Estimator Evaluation:}
The bias and standard deviation of the KSG and FMI estimator on Gaussian AR($N$) processes are shown in Fig. \ref{fig:voiced-a} for the orders 1 to 15. The corresponding autocorrelation function and the analytical mutual information are show in Fig. \ref{fig:mutualinformation} in the appendix. One set of AR(N) parameters were extracted from a sustained /a:/ vowel, while another was extracted from a sustained /s/. The different parameter sets were supposed to simulate the dependency structure of voiced (/a:/) and unvoiced (/s/) speech. The error bars in Fig. \ref{fig:voiced-a} (a) and (b) show 0.25 times the standard deviation $\sigma$ of the estimation.

\subsection{Maximum Prediction Gain of Sustained Speech:}
Fig. \ref{fig:three_phonemes} shows the prediction gain (PG) of all used predictors and upper bound estimators for two phonemes (/a:/ and /s). Analogous results are shown in the Appendix in Fig. \ref{fig:pg_e} for the /e/ phoneme. The depicted result for the /s/ phoneme is representative for all unvoiced phonemes, where the linear predictor was generally found to be optimal within about 0.3 dB and only the FMI PGMAXE suggested a slightly higher maximum PG, which can be explained with the observed FMI bias on AR($N$) processes modelling unvoiced speech as in Fig. \ref{fig:FMI_unvoiced}.
\begin{figure*}[h!] 
\centering                           
\subfloat[\label{fig:deltaPG_e}]{\includegraphics[width=0.49\textwidth]{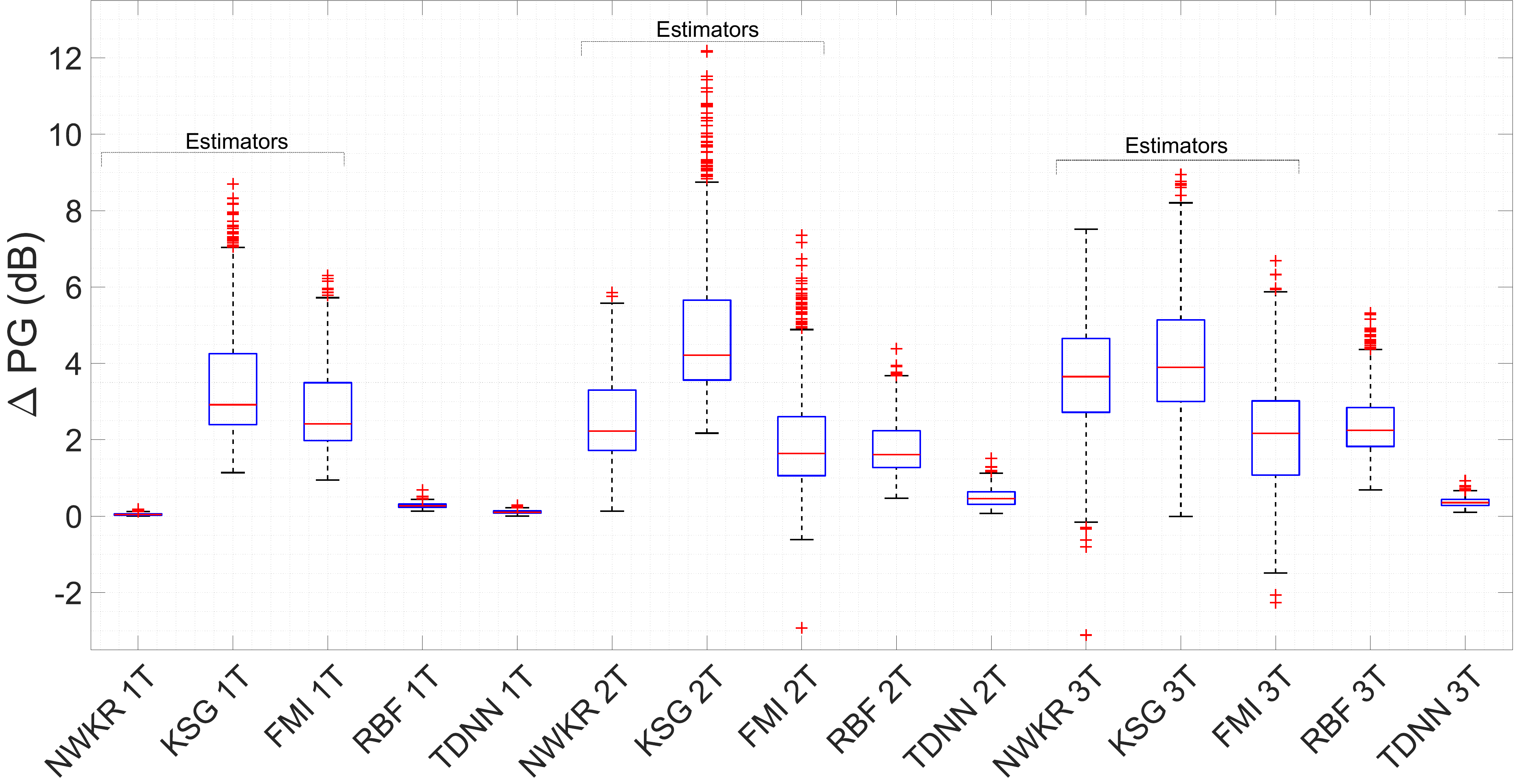}}\hfill
\subfloat[\label{fig:deltaPG_e_perSubject}]{\includegraphics[width=0.49\textwidth]{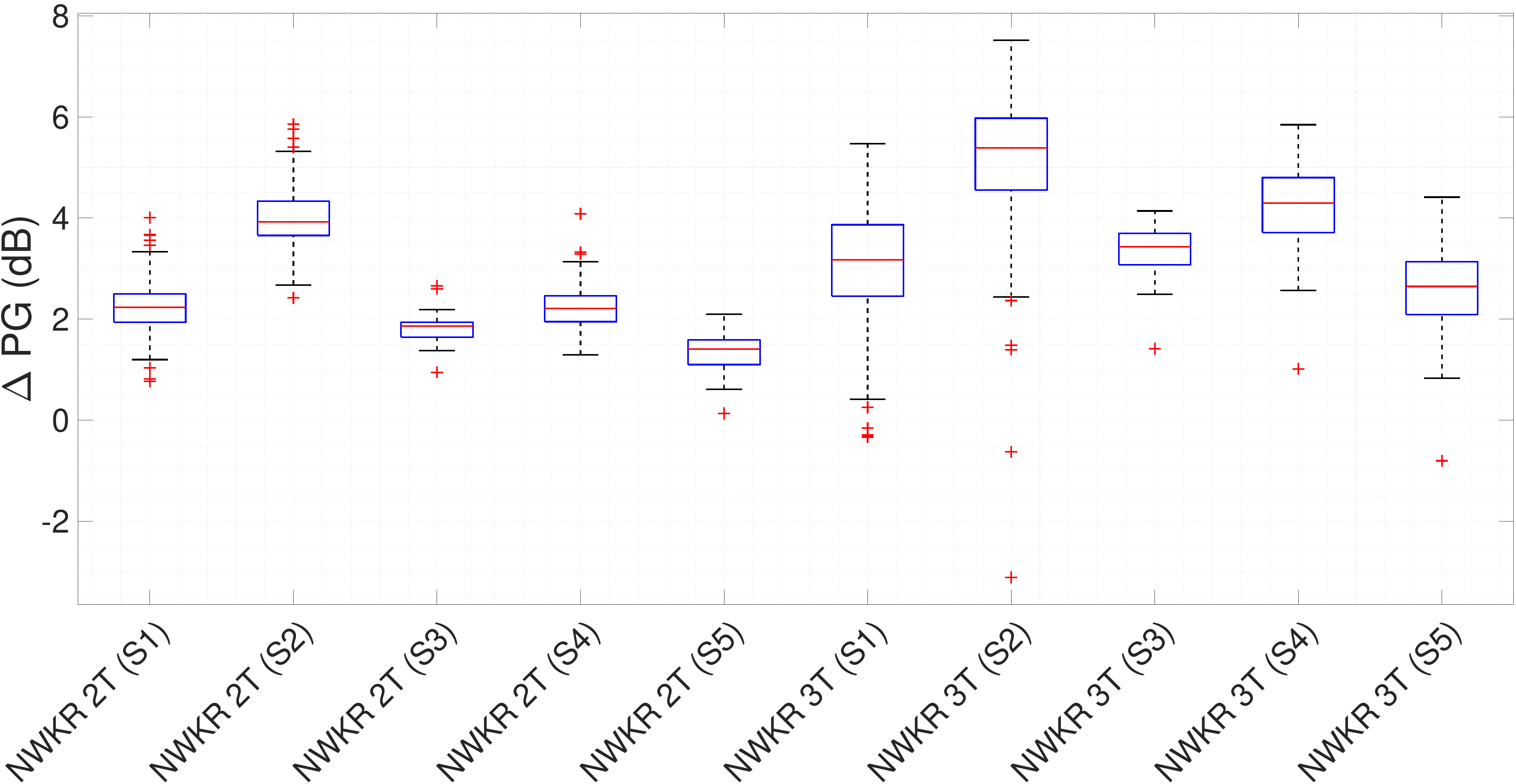}}  \\
\subfloat[\label{fig:deltaPG_NWKR_NN}]{\includegraphics[width=0.49\textwidth]{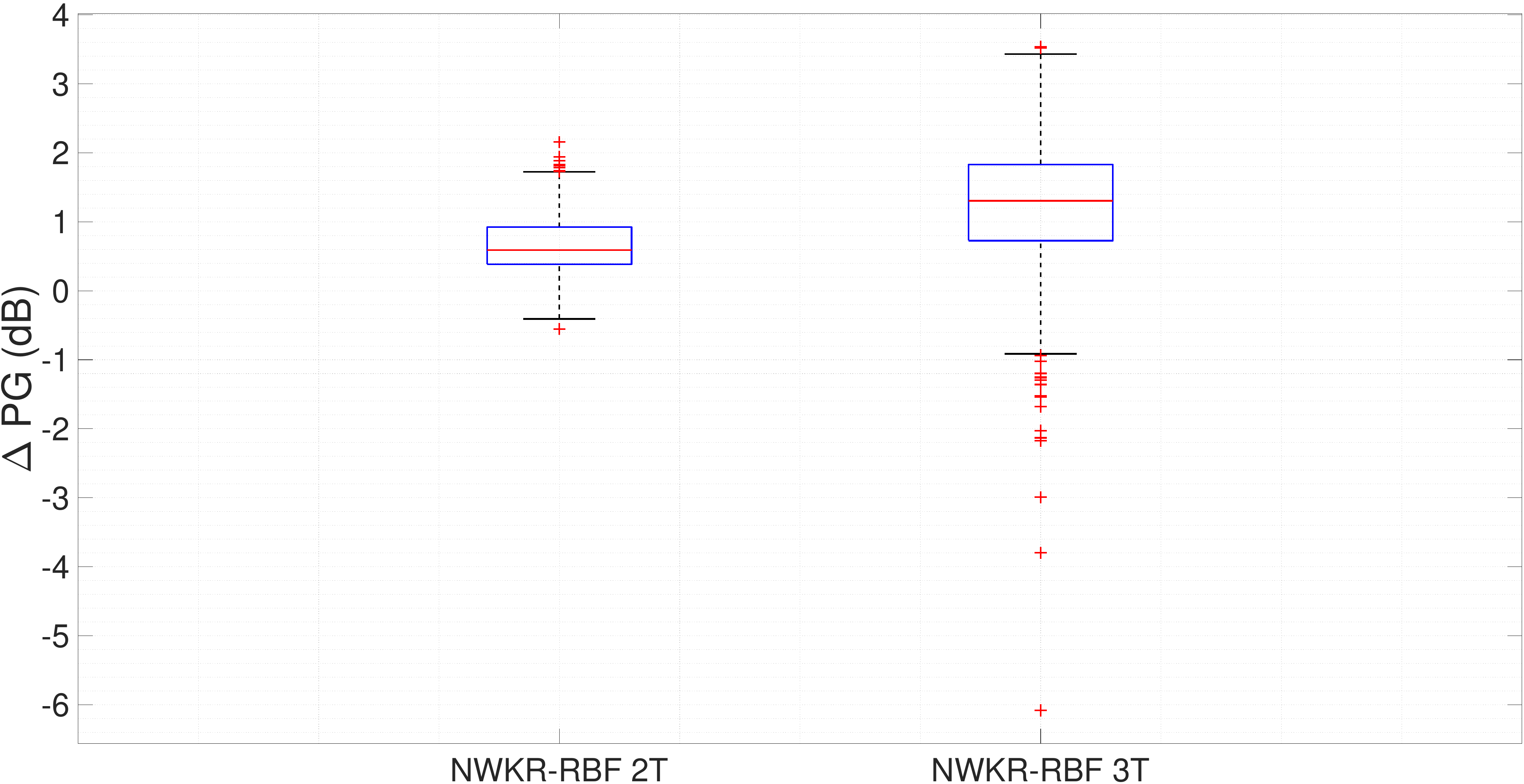}}\hfill
\subfloat[\label{fig:equiv_taps}]{\includegraphics[width=0.49\textwidth]{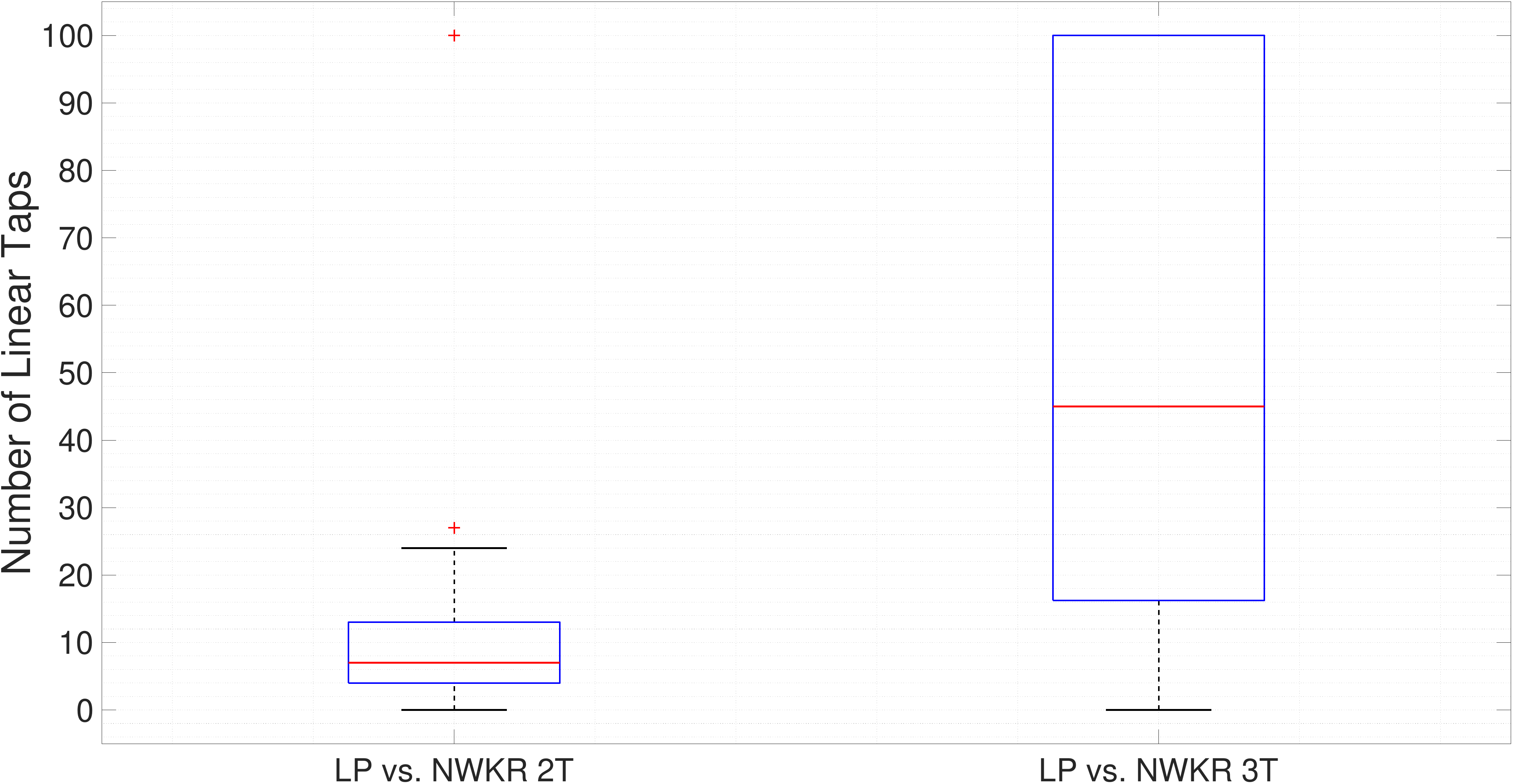}} 
\caption{\label{fig:e_results} (a) Difference $\Delta PG = PG_{NL} - PG_{Lin}$ of the $PG_{NL}$ of the investigated nonlinear methods and the static linear predictor $PG_{Lin}$ assessed for the /e:/ phoneme for one tap (1T) to three taps (3T), (b) $\Delta PG = PG_{NL} - PG_{Lin}$ for NWKR 1T to NWKR 3T assessed for the /e:/ phoneme across subjects, (c) $\Delta PG = PG_{NWKR} - PG_{RBF}$ between the PG of the NWKR and the RBF and (d) number of taps required by a static linear predictor (LP) to obtain the same PG as achieved by the NWKR 2T and NWKR 3T, respectively. The search was limited to a maximum of 100 taps.
 }
\end{figure*}
Fig. \ref{fig:deltaPG_e} shows a comparison of all predictors and all estimators across all segments of all subjects for the /e:/ phoneme. Only the best training result for the RBF and TDNN was used per segment. All subjects achieved at least a few stationary segments for this phoneme. Depicted is the difference $\Delta PG := PG_{NL}-PG_{Lin}$ in dB, where $PG_{NL}$ is the  achieved or estimated PG by the respective nonlinear method and $PG_{lin}$ is the PG of the static linear predictor with the respective tap number. The red bars denote the median performance, the bottom and top edges of the boxes mark the 25\,\% and 75\,\% percentile, respectively. The whiskers denote the most extreme points not considered outliers and the red marks denote outliers, where 1.5 times the interquartile range was used as threshold. Fig. \ref{fig:deltaPG_e_perSubject} shows $\Delta PG$ in dB, defined as above, across subjects S1 to S5 for two and three taps of the NWKR. 

Fig. \ref{fig:deltaPG_NWKR_NN} depicts the difference $\Delta PG$  in dB between the PG of the NWKR and the RBF, denoted as $NWKR-RBF$, for two and three taps.

Fig. \ref{fig:equiv_taps} depicts the number of taps that a static linear predictor requires to achieve the same PG as the NWKR on a given segment.  Segments, where the NWKR achieved lower PG than the linear predictor were omitted as these results are due to unfortunate optimization results. The maximum considered tap size was 100, i.e. if the linear predictor had not yet achieved the same PG as the NWKR with 100 taps, the equivalent tap size was set to 100. Tests showed that in some cases, even a 500 taps linear predictor did not achieve the performance of the three taps NWKR.

\begin{figure*}[t] 
\centering                           
\subfloat[\label{fig:i_deltaPG}]{\includegraphics[width=0.49\textwidth]{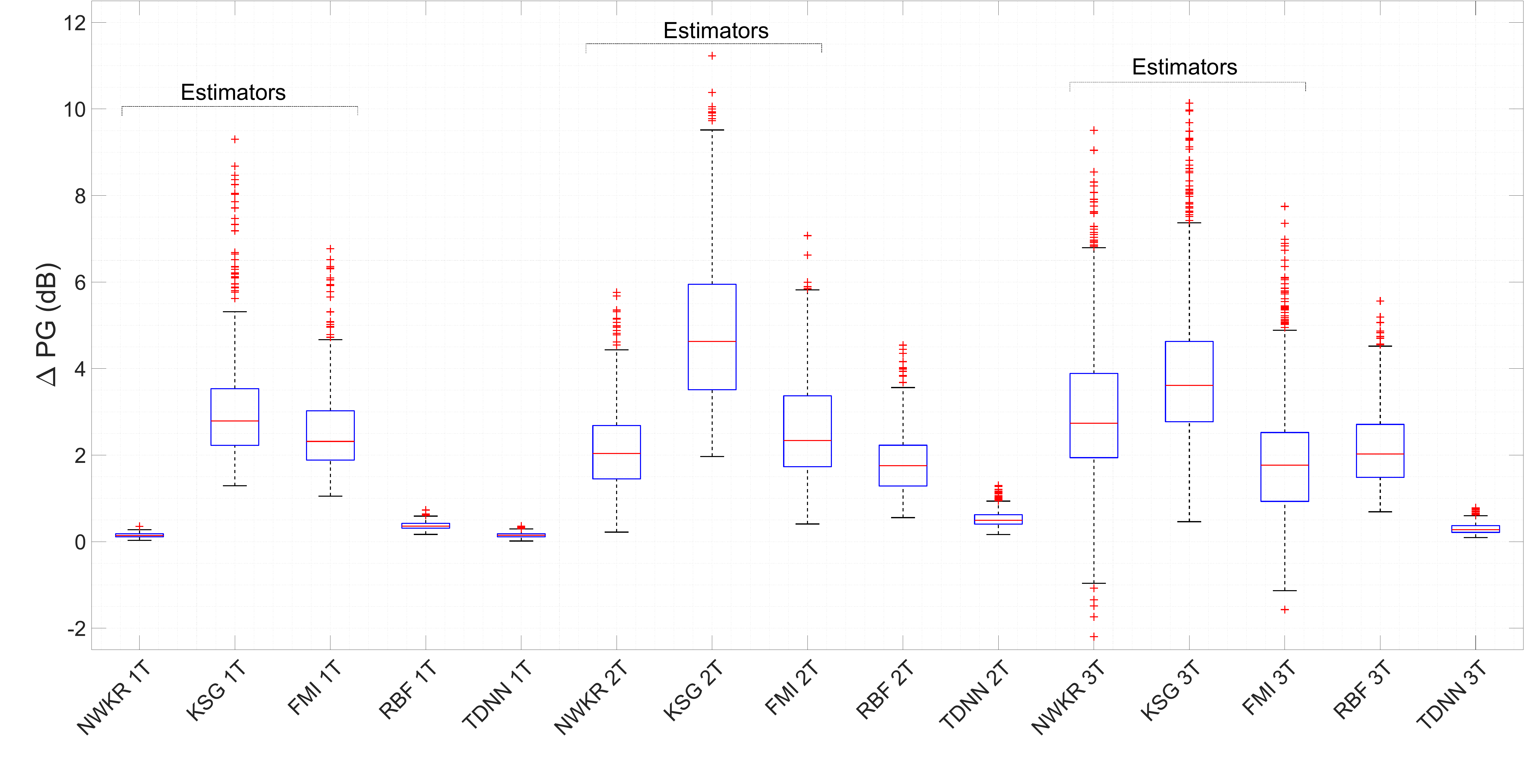}}\hfill
\subfloat[\label{fig:a_deltaPG}]{\includegraphics[width=0.49\textwidth]{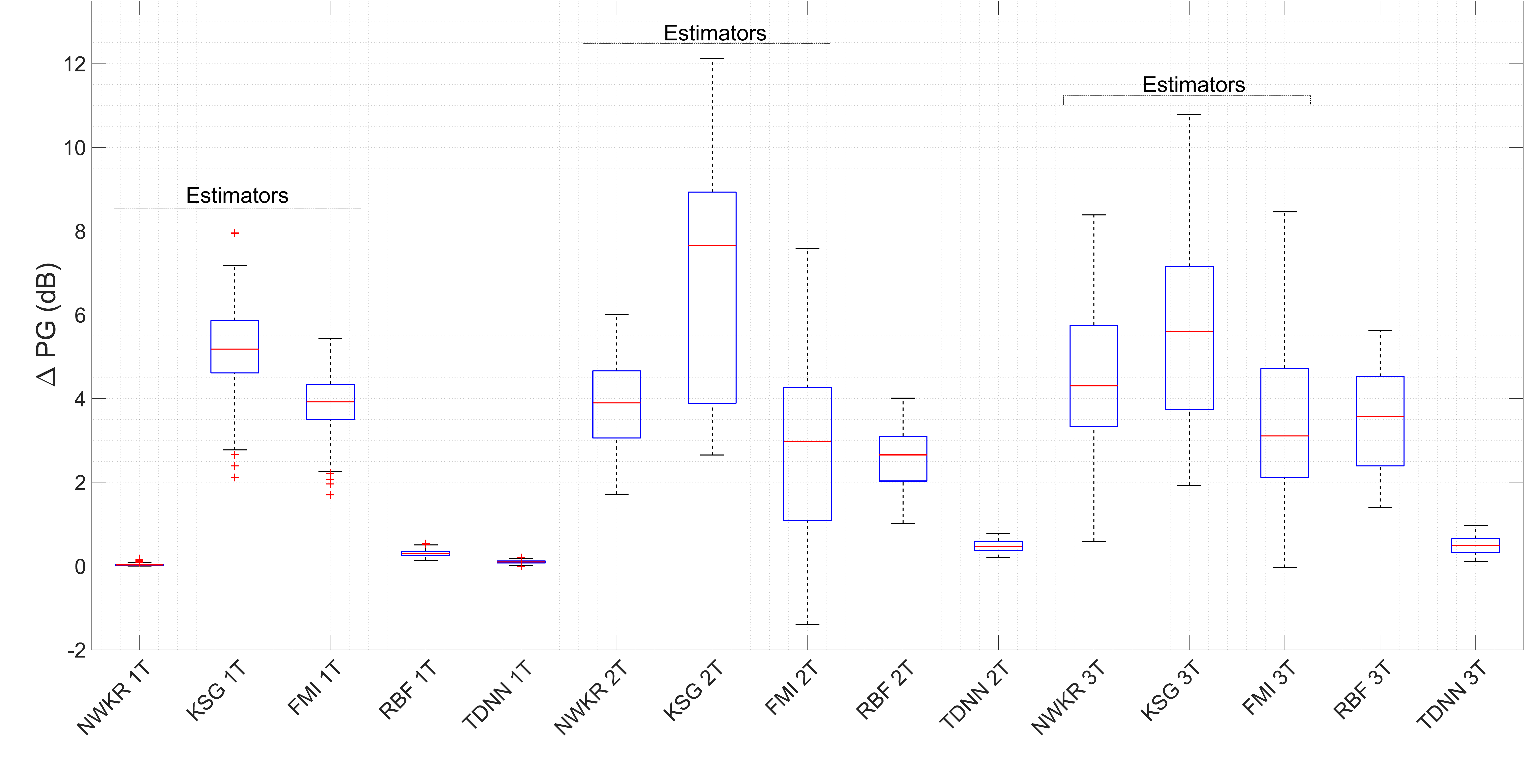}} 
\caption{ (a) Difference $\Delta PG = PG_{NL} - PG_{Lin}$ for the /i:/  and (b)  for the /a:/ phoneme for all estimators and all predictor models across all subjects. $PG_{Lin}$ is the prediction gain of the static linear predictor. $PG_{NL}$ is the prediction gain of the respective nonlinear methods (estimator or predictor). \label{fig:a_i_deltaPG}
 }
\end{figure*}

Fig. \ref{fig:a_i_deltaPG} shows a comparison of all predictors and all estimators across all segments of all subjects for the /i:/ (Fig. \ref{fig:i_deltaPG}) and /a:/ (Fig. \ref{fig:a_deltaPG}) phoneme. Note that the stationary /a:/ and /i:/ segments were mostly obtained from two subjects each, i.e. they only represent a subset of the subjects. The apparently higher maximum PG of the /a:/ phoneme is mostly due to subject S2, for which also the highest maximum PG was observed for the /e:/ phoneme.

\begin{figure*}[b]
\minipage{0.49\textwidth}
  \includegraphics[width=\linewidth]{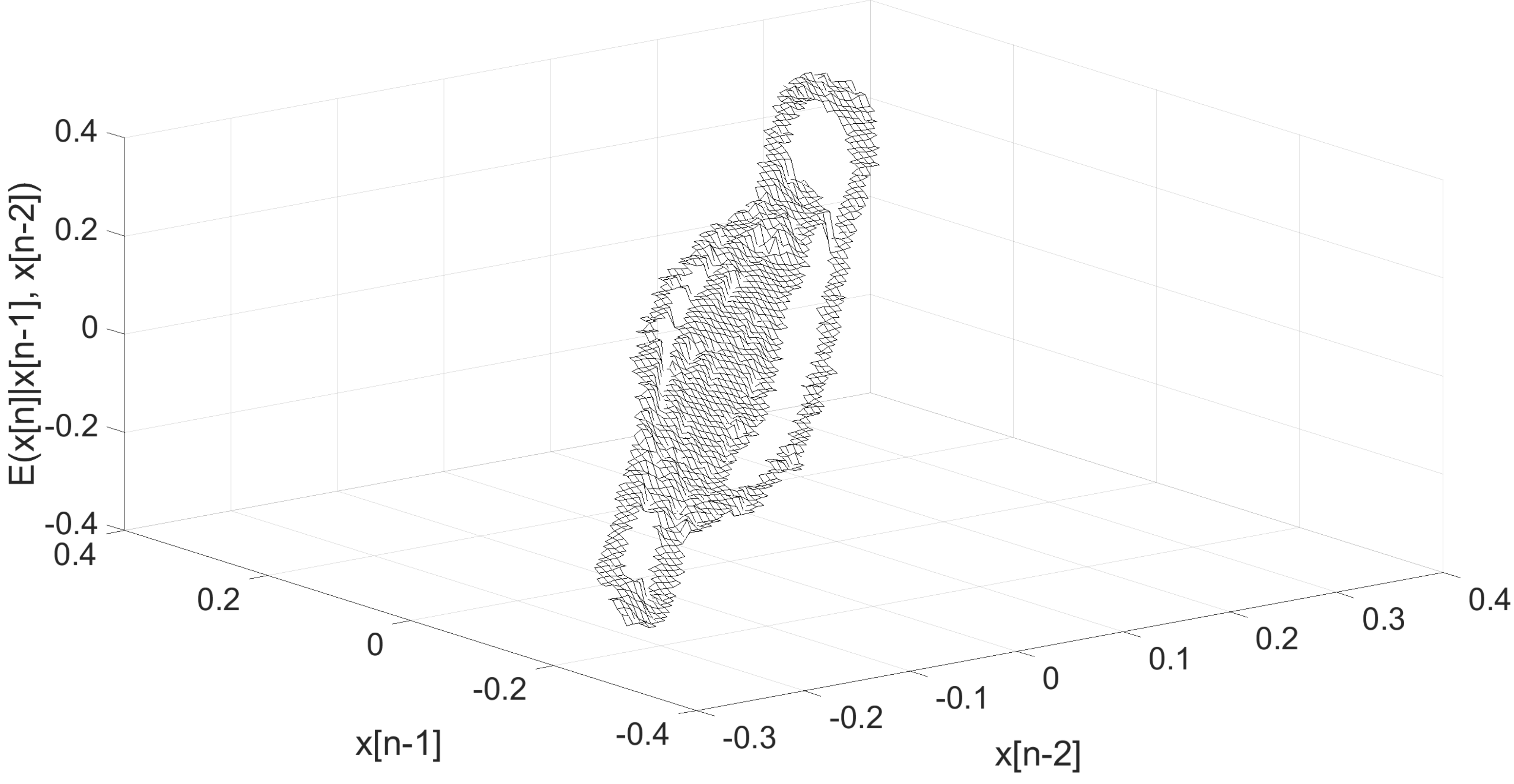}
\endminipage\hfill
\minipage{0.49\textwidth}%
  \includegraphics[width=\linewidth]{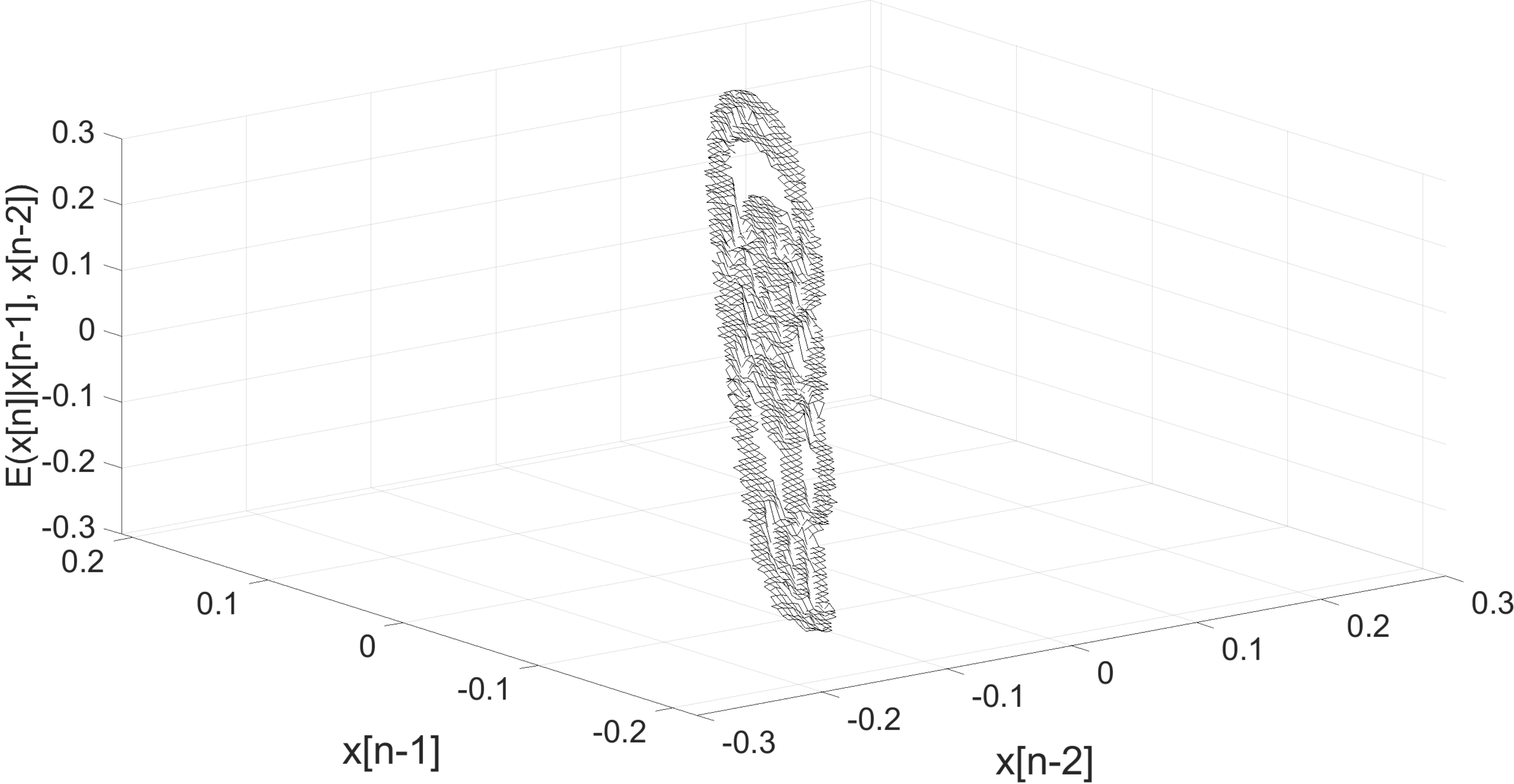}
\endminipage
  \caption{Conditional expectation $\hat{E}(x[n]|x[n-1], x[n-2])$ for one /a:/ segment (left) and one /i:/ segment (right). Both predictors achieved an about 6 dB higher PG than the  linear predictor and a more than 1.5 dB higher PG than the RBF.}\label{fig:awesome_image3}
\label{fig:cond_exp_a_and_i}
\end{figure*}

Fig. \ref{fig:cond_exp_a_and_i} shows two estimates of the conditional expectation as obtained with the NWKR and a 100 x 100 grid, ranging between 1.05 times the minimum and maximum value of the respective segments. Both obtained on their segment about 6 dB higher PG than the static linear predictor and at least 1.5 dB higher PG than the RBF.

\section{Discussion}
\label{sec:discussion}
This work shows that it appears to be possible to sustain speech sufficiently stationary to estimate the underlying conditional distribution, albeit it can be difficult as for /n/, /ng/, /o:/ and /u:/ no subject achieved stationary speech production. The forerunner experiment in\CITE{bernhardPhD} did not verify or report the stationarity of the recordings, therefore this result is reassuring. Our findings confirm the longstanding and wide-spread belief in speech prediction that unvoiced speech can be (nearly) optimally predicted by a linear predictor. The observed median difference of at most about 0.3 dB to the maximum prediction gain (PG) estimate made further analysis unnecessary. For voiced speech, the PGMAXE and NWKR estimate both indicate possible (median) improvements over the linear predictor of at least 2 dB to 3.5 dB depending on the number of taps and phoneme. Going by the NWKR results, in some cases an improvement of around 8 dB appears possible, which would represent a substantial gain over linear prediction. 

The median values agree well with experimentally obtained prediction gains in\CITE{Zhao18}, where an (adaptive) echo state network obtained an about 2.5\,dB -- 3\,dB higher prediction gain than an adaptive linear predictor using three taps each.
In contrast, the linear predictor achieved the maximum PG in\CITE{bernhardPhD}, but for a different so called reconstruction delay, which is therefore not comparable. The maximum PG appeared to be significantly higher for the /a:/ than for /e:/ and /i:/ phonemes. However, the stationary /a:/ segments stemmed mainly  from subject S2, which also showed significantly higher maximum PG for the /e:/ phoneme. This could indicate a difference between subjects, but care has to be taken in the interpretation, as the vocal properties  -- e.g. the amount of twang\CITE{Twang} -- of the subjects were not entirely identical. How such vocal properties affect the dependency structure of speech is entirely unknown.

The gaps in the conditional expectations shown in Fig. \ref{fig:cond_exp_a_and_i} occur around  turning points of the waveform, where the data might be sparser. Generally, the results of the PGMAXE and the NWKR for three taps has to be cautiously interpreted.   From the results, the PGMAXE, which in general does not represent a tight, achievable upper bound, might substantially overestimate the maximum PG, as the suggested median improvements over the one tap static linear predictor of up to 5.5 dB seem unreasonably large. Also, for three taps, the NWKR, KSG and FMI might be affected by the curse of dimensionality. The increased number of outliers in Fig. \ref{fig:deltaPG_NWKR_NN} for three taps could be related to this. From Fig. \ref{fig:KSG_voiced}, a negative bias of at least 0.75 dB for the KSG and 2.4 dB for the FMI is to be expected. 

The RBF performed about 0.5 dB to 1.5 dB within the lower estimates of the (median) maximum PG, achieving similar ($\pm 0.3 \, \text{dB}$) performance throughout the ten repetitions. The RBF obtained 0.6 dB and 1.3 dB lower median PG for two and three taps, respectively, than the NWKR and the largest differences amount to about 2 dB and 3.4 dB for two and three taps, respectively. For one tap, the RBF outperformed the linear predictor by about 0.35 dB median PG, perhaps indicating a slight overfit. The TDNN, in contrast to\CITE{thyssen1994non}, performed only slightly better than the linear predictor. However, in\CITE{thyssen1994non} a sampling rate of 8 kHz and more taps  were used. The observed gains in the PG of the applied nonlinear methods and predictors is on average lower than in\CITE{Birgmeier96}, but the sampling rates are not identical and\CITE{Birgmeier96} did not control the stationarity of the signals.

Our results suggest that, due to the observed intersubject differences in the estimated maximum PG, care has perhaps to be taken when comparing nonlinear predictors on different datasets.
With the presented approach and using the NWKR, the optimum predictor can in principle be obtained. This could be used to develope novel, low complexity predictor models in the future.

\subsection{Estimator Comparison}
In general the three different methods do not yield identical upper bounds of the prediction gain. The NWKR without a doubt yields the best estimate, and if anything, slightly underestimates the true upper bound. In all tests the NWKR always achieved the true maximum prediction gain or achieved slightly smaller values. In some cases the optimization fails which can result in pretty bad prediction gains which are obviously noticed.

There are two main reasons for the difference between the maximum prediction gain estimates of the KSG and FMI and the NWKR: 
The FMI, and to a lesser degree, the KSG, suffer from bias according to our mutual information investigation. From the AR($N$) simulations the bias of the FMI’s mutual information estimate could result in a bias in the bound of the maximum prediction gain be as large as -2.4 dB, that is, the actual upper bound could be 2.4 dB higher. The bias is smaller for the KSG, about 0.9 dB. The variance is rather small for the KSG and a little bit larger for the FMI, but the main issue is the bias. The bias can be reduced if longer segments could be used.
A second reason, more relevant than the bias, is the following observations we made after observing the differences between the NWKR estimate and the KSG/FMI estimate: 
The  tight upper bound  does not appear to be truely tight. It is tight in the case of Gaussian processes. But the Delta quantity in the definition of the upper bound by Bernhard is defined as the difference between the entropy of a Gaussian variable (with variance equal to the signal to be predicted) and the entropy of the signal to be predicted. 
Now for, e.g., independent Laplacian noise this entropy difference should be larger than zero (as the normal distribution has the maximum entropy property), but certainly the conditional expectation would be zero, so the best predictor would predict zero, and thus not change the variance of the signal. 
Accordingly, the actual maximum prediction gain in this example should be zero, yet this upper bound yields some value greater than zero. How large this gap is or can be is unfortunately unknown.

If a predictor achieves a prediction gain less than the NWKR it can not be optimal.
Due to the explained apparent gap between the upper bound by Bernhard, which is computed based on the KSG/FMI estimators, a predictor will never surpass their prediction gain. Thus the NWKR bounds the true maximum prediction gain from below (by experience, the difference should be rather small on average) while the KSG/FMI estimations will bound it from above.

\subsection{Practical Implications}
Predictor models have at least two broad applications, namely prediction
and  modeling, i.e., speech synthesis, as in LPC and similar applications.

Unlike many other domains of engineering, there is very little information on how good predictors can be in predicting (voiced) speech signals. Where in other domains the performance in terms of, e.g., efficiency of, for example, an electrical engine can be compared to theoretical bounds derived via physical modeling\CITE{Roshandel23}, there is almost no such thing for speech prediction. Instead different predictor models or ways to optimize them have been extensively tested on different datasets that contain natural speech. 
Generally, in speech coding, speech is classically distinguished\CITE{Kondoz05} in unvoiced and voiced segments, typically separated by brief transitional parts. The transitions are a separate problem that is difficult to investigated.
While unvoiced segments have been known for decades to be nearly optimally predictable (and thus also modelable) by linear predictors/models, voiced segments have been known for decades to contain nonlinear dependencies. The structure of these nonlinear dependencies, however, remains unknown. 
To shed light on these, more data – stationary data – is needed to not only be able to determine how good a predictor can become, but also to be able to actually devise new predictor models that allow to achieve the maximum possible prediction gain. Then it is certain that the prediction cannot be improved.
So far a large number of different number of methods to nonlinear prediction has been applied in the literature on all kinds of datasets with varying performance gains over linear prediction\CITE{Birgmeier97,Zhao18,Volterra}.
Now, even if the most promising approaches (likely some sort of neural networks) were to be compared to each other on the same dataset, one would not know whether the best performing model was actually achieving the best possible prediction gain or whether it was stuck in a local optimum. Also, for resource constrained devices like hearing aids, smaller predictor models are particularly beneficial\CITE{TinyLSTM}.
As such, knowledge of the functional form of the optimal predictor can be advantageous.

This situation could be improved by the approach of our paper:  Sampling (voiced) speech sufficiently long to not only get estimates of the maximum achievable prediction gain, but in the case of the NWKR, to obtain, through kernel density estimation, the conditional expectation that is known to be the best predictor with respect to the mean-squared error. Then one can also directly compare the output of other predictor models to the conditional expectation and use, e.g., symbolic regression to attempt to identify the function implicitly given by the NWKR.

Although 10,000 samples appear to be a sufficient amount to obtain reasonable estimations, some uncertainties are left and to be investigated in the future. Selecting "experts" in sustained phoneme production, perhaps professional singers or voice actors, one could potentially vastly increase the length of the stationary segments. Another way would be to record substantially more repetitions. Sometimes, longer segments were found to be sufficiently stationary. That way, the estimations of the maximum prediction gain could be further improved.

This work investigated the maximum prediction gain on sustained speech. Natural speech does not contain extensively sustained phonemes.
To tie this back to natural speech, note that several (natural) speech coding methods, most prominently LPC and code-excited linear prediction (CELP) use speech models to synthesis or code speech signals. CELP for example switches between voiced and unvoiced models on a per frame basis. 
CELP models voiced segments by using previous segments to generate excitation noise to drive a linear predictor/model\CITE{Spanias94} (adaptive codebook). If a novel predictor model for voiced speech was developed, which could be done using our method, perhaps CELP-like approaches could be improved. The same applies to LPC-based algorithms which also modern neural audio codecs like LPCNet\CITE{LPCNet} make use of. LPC does not switch prediction models but always uses a linear predictor. Again, the model potentially could be improved with a superior predictor model. This would have an immediate practical application.

Finally, knowledge on the functional form of the conditional expectation, i.e., the optimal predictor of speech, could potentially help in understanding neural networks (in the context of interpretable machine learning/neural networks\CITE{Zhang21}).

\subsection{Future Work}
Novel nonlinear prediction models could have relevant applications in ultra low latency, low resource settings like wireless transmission of audio in hearing aids but also in neural audio codecs like LPCNet.

The approach of this paper --  sustaining phonemes, checking their stationarity followed by kernel regression -- could be used to develop such predictor models.
Although the NWKR does not yield an explicit predictor function, it could perhaps be used in symbolic regression as a target to find a closed form expression.
For this purpose, a cohort of "expert" subjects should be invited with great control over their vocal tract. Potential candidates would be professional singers or voice actors. They could potentially allow to record considerably longer stationary segments so that the estimation bias of the mutual information estimation, but also the NWKR bandwidth optimization, can be made very small.

Also, long term prediction was not investigated in this work. Future work could expand the proposed method to include long term prediction of speech which is known to allow to considerably improve voiced speech prediction.

\section{Conclusion}
This work investigates the maximum achievable prediction gain (PG) on sustained speech. Linear and nonlinear predictors including a radial basis function network (RBF) were compared to two independent estimates of the maximum PG. For unvoiced speech, the linear predictor was found to be optimal within about 0.3 dB of the estimated maximum PG. For voiced speech, a possible median improvement of between 2 dB and 3.4 dB depending on tap number and subject was found. A median number of seven and 45 taps was found to be required for the linear predictor to match the estimated two tap and three tap maximum PG, respectively.  The median PG of the RBF was found to be suboptimal by about 0.6 dB to 1.3 dB.

\section*{Acknowledgements}
We would like to thank Hans-Peter Bernhard for his support and for providing us with his original code. Furthermore, we thank all subjects of our study.

\section*{Funding}
The research has not been funded by third parties. 


\section*{Availability of data and materials}
All recordings and the code can be received upon request for research purpose only.

\section*{Ethics approval and consent to participate}
All subjects gave their informed consent to participate. All subjects agreed to anonymous usage of their results.

\section*{Competing interests}
The authors declare that they have no competing interests.

\section*{Consent for publication}
All subjects gave their informed consent to anonymous usage of their data and results in scientific publications.

\section*{Authors' contributions}
R.H.: Idea, main author, implementation of the NWKR, data analysis\\
M.D.: Analysis of KSG and FMI estimators, neural network Matlab programming  \\
S.P.: Recording, Proof-reading\\
J.O.: Supervision, Proof-reading\\
R.H. and M.D. declare equal contribution.


\newpage 
\section*{Appendix}
\begin{figure*}[h] 
\centering                           
\subfloat[ACF (Voiced and Unvoiced)]{\includegraphics[width=0.49\textwidth]{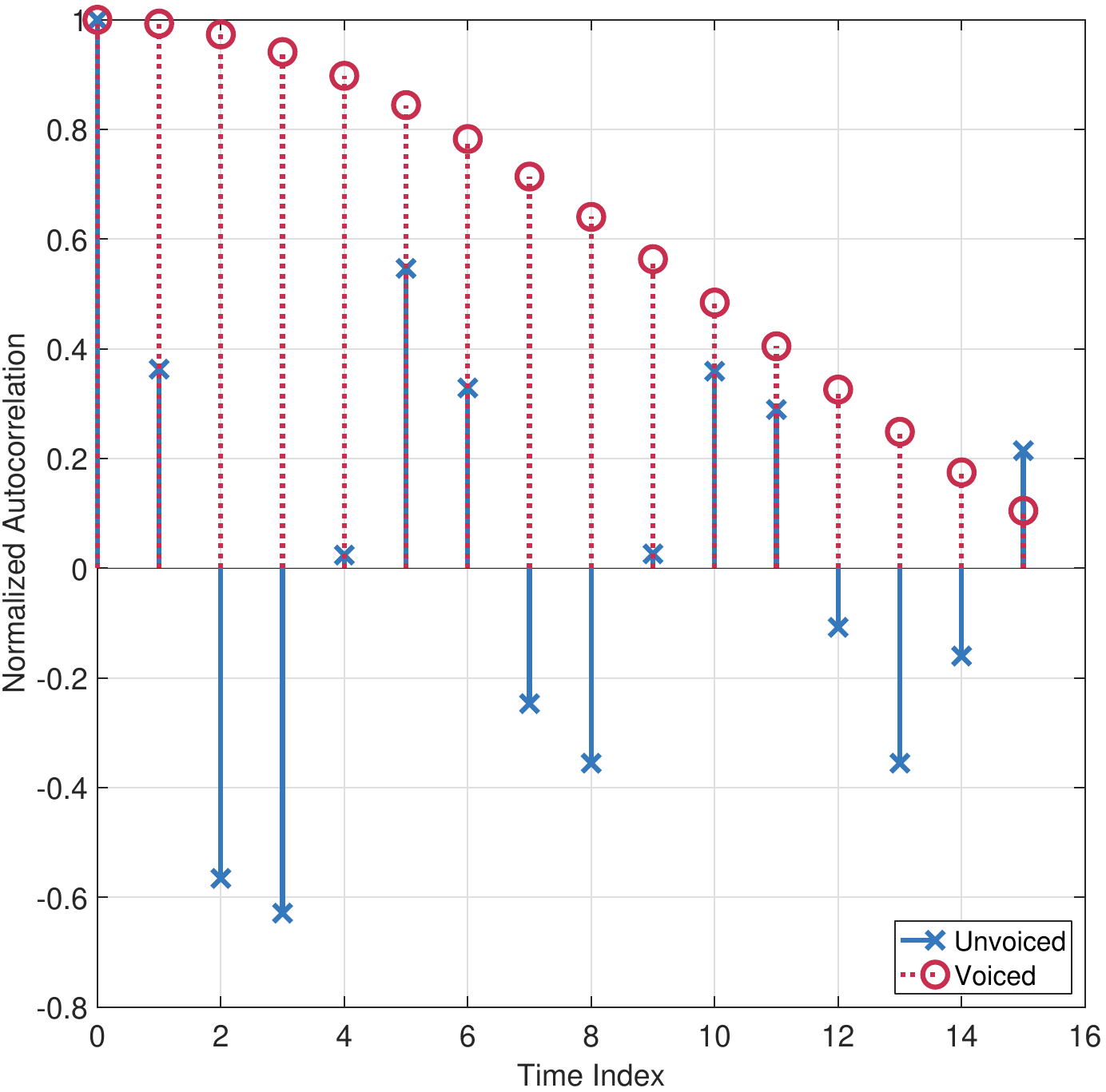}} 
\subfloat[Mutual Information]{\includegraphics[width=0.49\textwidth]{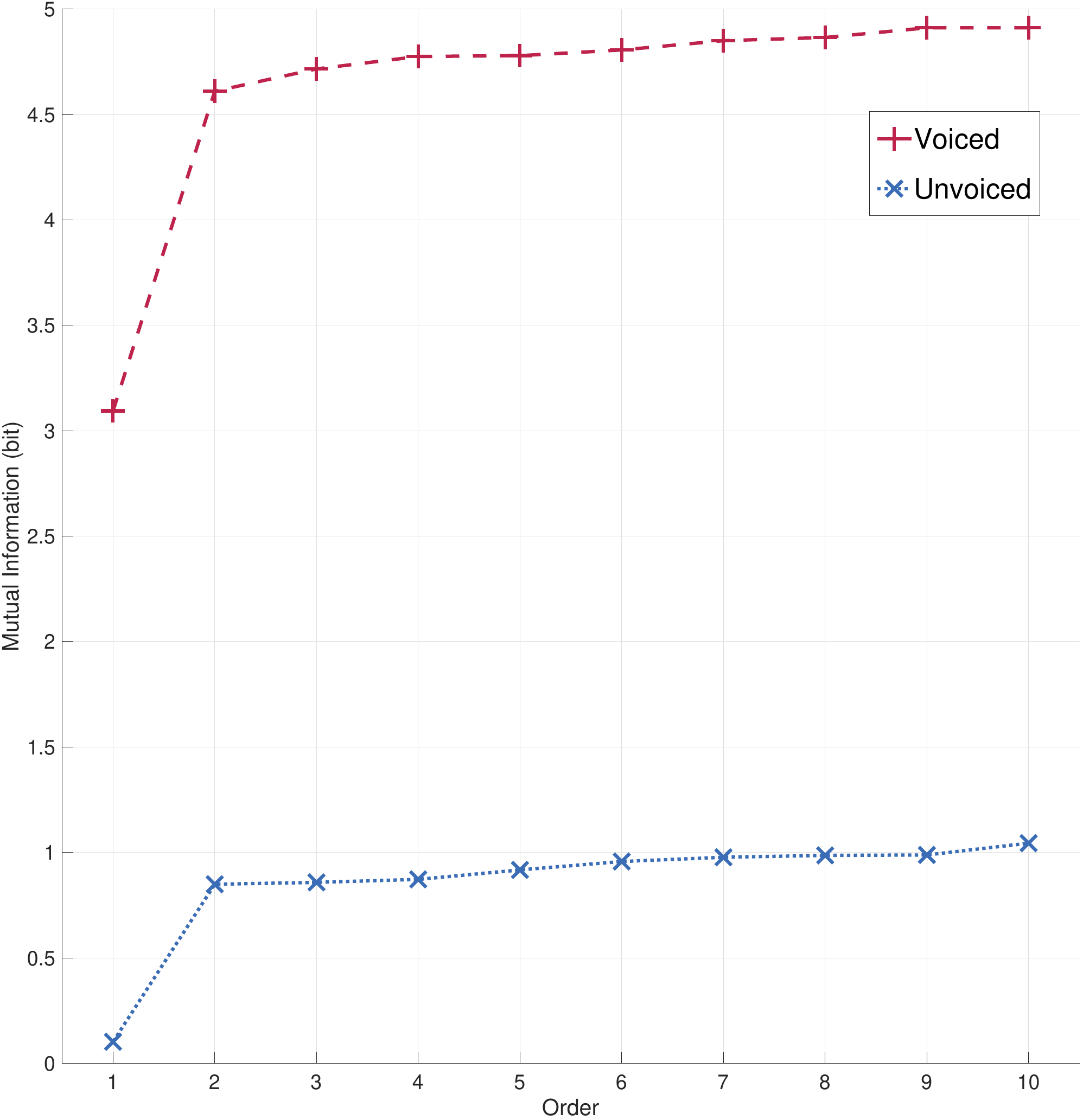}} \\
\caption{The autocorrelation functions (ACF) are depicted in (a). The analytical mutual information (MI) according to Eq. \ref{eq:Igauss} is shown in (b). 
 }
 \label{fig:mutualinformation}
\end{figure*}
\begin{figure*}[h!]
     \includegraphics[width=\linewidth]{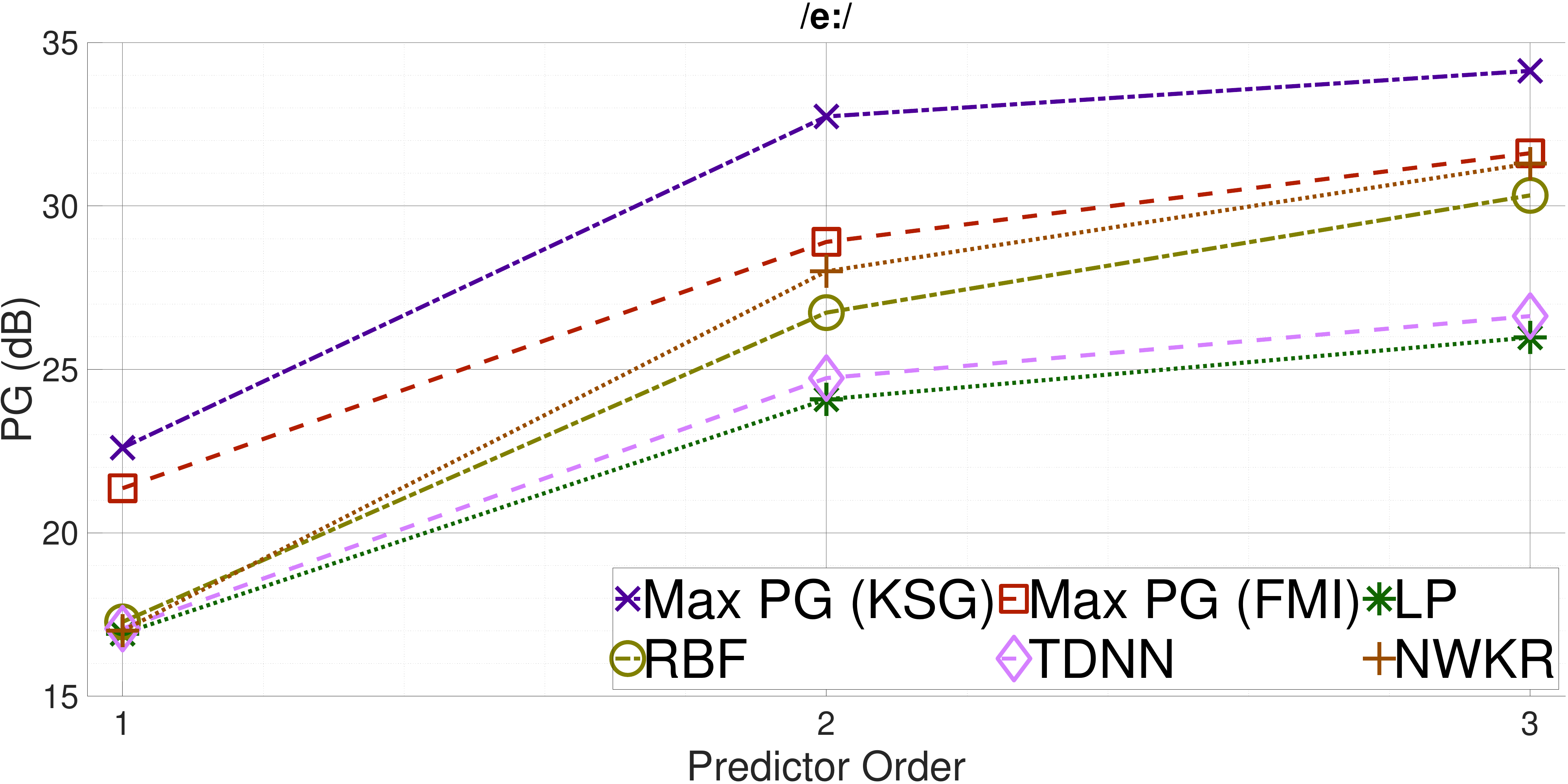}
     \caption{Prediction gain for all predictors and all estimators of the maximum prediction gain for the /e:/ segment produced by a male subject.}
     \label{fig:pg_e}
\end{figure*}
\minipage{0.32\textwidth}
\endminipage\hfill

\end{document}